\documentclass[twocolumn,aps,superscriptaddress,showpacs,amsmath,amssymb]{revtex4-2}

\usepackage{graphics}
\usepackage{graphicx}
\usepackage{dcolumn}
\usepackage{bm}
\usepackage{color}
\usepackage{soul}
\usepackage{textcomp}
\usepackage{multirow}
\usepackage{setspace}
\usepackage[mathlines]{lineno}
\usepackage{textcomp}
\usepackage{mathcomp}
\usepackage{spverbatim}
\usepackage{float}
\usepackage{longtable}

\usepackage{xcolor}

\usepackage{hyperref}
\hypersetup{
    colorlinks=true,       
    linkcolor=blue,        
    citecolor=blue,        
    filecolor=blue,        
    urlcolor=blue,         
    runcolor=blue
}

\newcommand{\ket}[1]{\left| #1 \right\rangle}

\newcommand{\meanval}[1]{\left\langle #1 \right\rangle}

\draft 

\begin{document}

\title{Ultrafast modulation of vibrational polaritons for controlling the quantum field statistics at mid-infrared frequencies}

\author{Johan F. Triana}\email{johan.triana@usach.cl}
\affiliation{Department of Physics, Universidad de Santiago de Chile, Av. Ecuador 3493, Estaci\'on Central, Santiago, Chile}

\author{Felipe Herrera}\email{felipe.herrera.u@usach.cl}
\affiliation{Department of Physics, Universidad de Santiago de Chile, Av. Ecuador 3493, Estaci\'on Central, Santiago, Chile}
\affiliation{ANID - Millennium Science Initiative Program, Millennium Institute for Research in Optics, Chile}

\date{\today}

\begin{abstract}
Controlling the quantum field statistics of confined light is a long-standing goal in integrated photonics. We show that by coupling molecular vibrations with a confined mid-infrared cavity vacuum, the photocount and quadrature field statistics of the cavity field can be reversibly manipulated over sub-picosecond timescales. The mechanism involves changing the cavity resonance frequency through a modulation of the dielectric response of the cavity materials using femtosecond UV pulses. For a single anharmonic molecular vibration in an infrared cavity under ultrastrong coupling conditions, the pulsed modulation of the cavity frequency can adiabatically produce mid-infrared light that is simultaneously sub-Poissonian and quadrature squeezed, depending on the dipolar behavior of the vibrational mode. For a vibration-cavity system in strong coupling, non-adiabatic polariton excitations can be produced after the frequency modulation pulse is over, when the system is initially prepared in the lower polariton state. We propose design principles for the generation of mid-infrared quantum light by analyzing the dependence of the cavity field statistics on the shape of the electric dipole function of the molecule, the cavity detuning at the modulation peak and the anharmonicity of the Morse potential. Feasible experimental implementations of the modulation scheme are suggested. This work paves the way for the development of molecule-based mid-infrared quantum optical devices at room temperature.
\end{abstract}

\maketitle 

\section{Introduction}

Quantum light sources are key to the success of optical quantum technologies \cite{Obrien2009}. 
In the visible spectral range, several mechanisms are available to produce quantum light, such as single photon emission by atoms in optical resonators \cite{Daiss2019,Agudelo2019}, photon blockade in cavity optomechanics \cite{Aspelmeyer2014}, or ensemble strong coupling with atomic emitters \cite{Saez2018}. In the near infrared  ($0.7 - 2.5 \,\mu{\rm m}$), quantum light generation is a mature technology that has become the workhorse for applications in quantum communication \cite{Kimble2008,Yu2015} and sensing \cite{Clark2021,Vega2020}. Quantum-efficient semiconductor \cite{Zhang2015} and superconductor \cite{Natarajan2012} photodetectors are broadly available in this frequency range. In contrast, quantum optical tools in the mid-infrared spectral region ($2.5 - 25 \,\mu{\rm m} $) are much less developed \cite{Shields2020,Spitz2021}. Quantum cascade lasers have been used to produce mid infrared quantum light \cite{Yao2012} and early demonstrations of quantum light detection in this frequency regime are available \cite{Gabbrielli2021,Mancinelli2017}.

In this work, we introduce a new mechanism for the deterministic generation and dynamical manipulation of quantum light in the mid-infrared frequency range. The mechanism involves ultrafast modulation of the confined mid infrared vacuum field under strong and ultrastrong light-matter coupling. Strong vibration-cavity coupling has been demonstrated with Fabry-Perot cavities \cite{Shalabney2015coherent,Shalabney2015raman,George2015,Long2015,Grafton2020,Xiang2019manipulating,Dunkelberger2016,Xiang2018,Dunkelberger2018}, plasmonic resonators \cite{Muller2018,Metzger2019},  van der Waals resonators \cite{Autore:2018} and polar dielectric resonators \cite{Folland:2020}. Ultrastrong coupling \cite{Kockum2019,Forn-Diaz2018} has also been demonstrated in the mid-infrared \cite{George2016,Askenazi2017,Yoo:2021}. The proposed modulation of the light-matter dynamics is based on the transient modification of the dielectric function of the materials that confine the infrared vacuum by driving interband transitions with a driving source. The associated change in boundary conditions for the electromagnetic field results in a parametric change of the cavity vacuum frequency that is related with the dynamical Casimir effect \cite{Dodonov2020fifty}. This frequency modulation mechanism has been demonstrated for mid-infrared resonators using electrical \cite{Jun2012}, chemical \cite{Panah2017} or electromagnetic \cite{Dunkelberger:2018,Dunkelberger2020tuning} driving sources.  

We consider a scheme where the mid infrared cavity frequency $\omega_{\rm c}\approx 1840 \,{\rm cm}^{-1}$ ($5.6\,\mu{\rm m}$) is modulated over sub-picosecond timescales by $200-400\, {\rm cm}^{-1}$ using a single femtosecond UV pulse that drives a target cavity boundary material to transiently modify its carrier density. We assume a nanophotonics scenario where the  intracavity medium has a single anharmonic molecular vibration in the ground electronic state, protected by design of the photonic structure from the  UV source. We  show that the photocount and quadrature field statistics of the infrared cavity field is transiently modified by the frequency modulation pulse and correlate the predicted changes with the intrinsic properties of the molecular vibration and the light-matter interaction process.

The article is organized as follows: In Sec. \ref{section:Theory}, we review the theoretical framework for cavity quantum electrodynamics with anharmonic molecular vibrations, and the numerically methods used for unitary wavepacket propagation. We also review the relevant concepts of quantum field statistics. We study the evolution of the intracavity field statistics for a Gaussian frequency modulation in vibrational ultrastrong coupling in Sec. \ref{sec:vusc} and discuss  strong coupling  in  Sec. \ref{sec:vsc}. We conclude and discuss possible experimental realizations in Sec. \ref{sec:conclu}.

\section{Methodology}
\label{section:Theory}

\subsection{MLQR model with cavity frequency modulation }
We model light-matter interaction of a confined cavity mode of frequency $\omega_{\rm c}$ with a single anharmonic vibration mode of fundamental frequency $\omega_{\rm v}$ in the mid-infrared as a Morse oscillator using the multi-level quantum Rabi (MLQR) model \cite{Hernandez2019,Triana2020}, recently developed to consistently describe the strong and ultrastrong vibration-cavity coupling regimes in the electric dipole approximation \cite{Andrews2018}. Taking into account the possible time-dependence of the cavity frequency due to external UV pulse driving, the corresponding MLQR Hamiltonian can be written in coordinate space as (in atomic units)
\begin{eqnarray}
\nonumber \hat{\mathcal{H}}(t)&=&\hat{H}_{\mathrm{M}}  -\frac{1}{2}\frac{\partial^{2}}{\partial\hat{x}^{2}} + \frac{1}{2}\omega_{\mathrm{c}}(t)^{2}\hat{x}^{2} + \sqrt{2\omega_{\mathrm{c}}(t)}\mathcal{E}_{0}(t)\hat{d}(q)\hat{x},  \\
\label{eq:hamiltonian}
\end{eqnarray}
where the vibrational Hamiltonian $\hat{H}_M$ includes the nuclear kinetic energy $\hat{T}(q)$ and the potential energy $\hat{V}(q)$ along the mass-weighted normal mode coordinate $q$. The second and third terms describe the quantized cavity oscillator with quadrature operator $\hat x$ and frequency $\omega_\mathrm{c}(t)$. The last term is the electric dipole light-matter coupling, proportional to the amplitude of the vacuum fluctuations at the cavity frequency $\mathcal{E}_{0}(t)=\lambda_{g}\omega_{\mathrm{c}}(t)/d_{10}$. 

The light-matter coupling strength is given in terms of the dimensionless parameter $\lambda_{g}$, which is equivalent to the conventional Rabi coupling ratio $g/\omega_{\mathrm{c}}$ \cite{Kockum2019,Forn-Diaz2018} for $g \equiv \langle 1|\hat d(q)|0\rangle\mathcal{E}_0 $, where $\ket{0}$ and $\ket{1}$ are the ground and first excited vibrational levels, $\hat{d}(q)$ is the electric dipole function \cite{Elsaesser1991} and $d_{10}=\langle1|\hat{d}(q)|0\rangle$. The onset of the conventional ultrastrong coupling regime is thus reached when $\lambda_g \sim 0.1$ \cite{Forn-Diaz2018}. Strong coupling implies $\lambda_g\gtrsim0.01$. 

Under UV pulse driving, the cavity frequency is taken to vary according to a Gaussian function as
\begin{equation}\label{eq:cavfreq}
\omega_{\mathrm{c}}(t)=\omega_{\mathrm{c}}\left( 1 + \eta e^{-(t-t_{\rm d})^{2}/2\tau^{2}}\right), 
\end{equation}
where the undriven cavity frequency being resonant with the fundamental vibration frequency ($\omega_{\mathrm{c}}(0)=\omega_{\rm v}$). As anticipated above, we assume the cavity frequency shifts at the UV pulse peak by up to 20\% ($\eta=\pm 0.2$), red-detuned (minus sign) or blue-detuned (plus sign) relative to the undriven resonance frequency. According to Eq. (\ref{eq:hamiltonian}), the frequency modulation not only changes the vibration-cavity detuning, but also the light-matter interaction strength. For the cases that will be discussed below, the peak modulation $|\eta|=0.2$ corresponds to changes in light-matter coupling energy of $12\%$ for $\lambda_g=0.08$ (strong coupling), and $30\%$ for $\lambda_g=0.2$ (ultrastrong coupling). Throughout this work, we assume the pulse is centered at $t_{\rm d}=250.0$ fs and the width parameter is $\tau= 62.5$ fs (FWHM = $147$ fs).

\begin{table}[b]
\caption{\label{tab:potentials} Parameters of Morse potential energy curves [Eq. (\ref{eq:morse})] and dipole moment functions [Eq. (\ref{eq:dipole})] employed in this work and plotted in Fig. \ref{fig:potentials}.}
\begin{ruledtabular}
\begin{tabular}{c c c c c}
 &          $D_{\mathrm{e}}$ [eV]     &         $\alpha$ [a.u.]         &  $q_{\mathrm{eq}}$ [bohr] & $\Delta_{21}$ [cm$^{-1}$] \\
 \hline
$V_{\mathrm{A}}(q)$  &  6.80  &  1.50  &  4.0  & 31.9\\
$V_{\mathrm{B}}(q)$  &  9.80  &  1.25  &  4.0  & 22.1 \\
\hline\hline
 &          $d_0$ [Debye]     &         $q_0$ [bohr]    &  $q_1$ [bohr] &  $\sigma$  bohr \\
 \hline
$d_{\mathrm{NP}}(q)$  &  5.08  &  4.0  &  4.0  &  0.6  \\
$d_{\mathrm{PR}}(q)$  &  2.54  &  2.7  &  4.5  &  0.6 
\end{tabular}
\end{ruledtabular}
\label{tab:morse parameters}
\end{table}

\begin{figure}[t]
\centering
\includegraphics[width=0.35\textwidth]{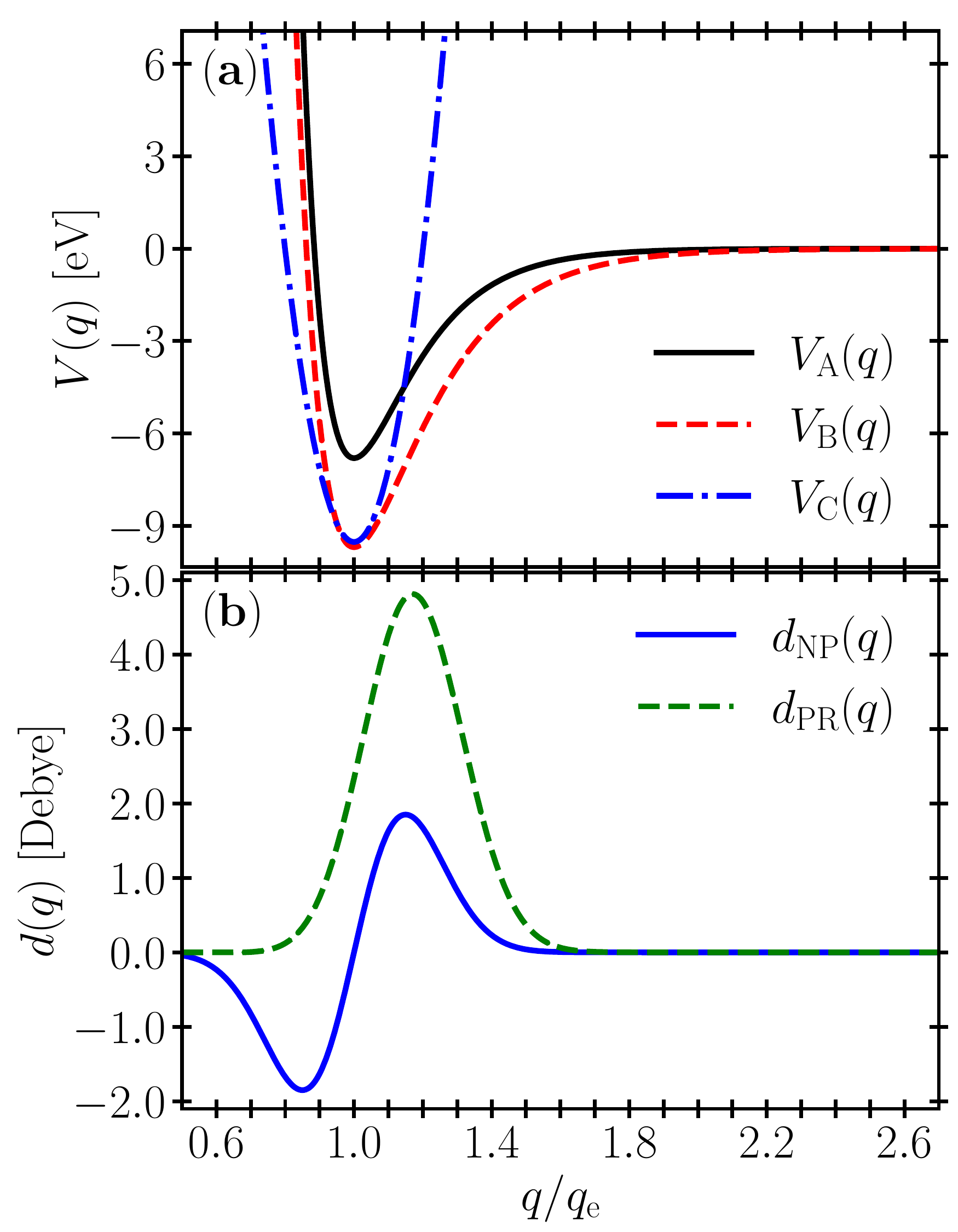}
\caption{{\bf Vibrational potentials and dipole functions}. (a) Morse potentials $V_\mathrm{A}$ (solid line) and $V_\mathrm{B}$ (dashed line) along the normal mode coordinate $q$. The harmonic oscillator potential $V_\mathrm{C}$ (dot-dashed line) with the same $\omega_{\rm v}$ is also shown. In all cases, the reduced mass is $\mu=8.5$ AMU and the fundamental frequency is $\omega_{\rm v}\approx1838$ cm$^{-1}$. (b) Dipole moment functions $d(q)$ for polar-right (dashed line) and a non-polar vibrational modes (solid line). }
\label{fig:potentials}
\end{figure}

\subsection{Dipolar Morse oscillator}

To understand the influence of the spectral anharmonicity and the dipolar structure of molecular vibrations on the unitary evolution of the frequency-modulated cavity field, we combine three different vibrational potentials $V(q)$ and two dipole moment functions $d(q)$ to search for conditions that are most effective at producing dramatic changes in the quantum field statistics of the infrared field. 

For anharmonic vibrations we use the Morse potential \cite{Morse1929}

\begin{equation}
V(q)=D_{\mathrm{e}}(1-e^{-\alpha(q-q_{\mathrm{eq}})})^{2}-D_{\mathrm{e}}, 
\label{eq:morse}
\end{equation}
where $D_\mathrm{e}$ is the classical dissociation energy, $q_\mathrm{eq}$ the equilibrium mode coordinate, and $\alpha$ the anharmonicity parameter. For $\alpha\ll 1$, the Morse potential can be accurately truncated at second order in $q$, giving the harmonic oscillator. In Table \ref{tab:morse parameters}, we list the parameters for the two Morse potentials shown in Fig. \ref{fig:potentials}a. To ensure a fair comparison between anharmonic and harmonic potentials, the parameters in Table \ref{tab:morse parameters} give the same vibrational frequency $\omega_{\rm v}=1838.26$ cm$^{-1}$, which is representative of the vibrational modes used for light-matter coupling in experiments \cite{Muller2018,Ribeiro2018,Grafton2021}. The harmonic potential $V_\mathrm{C}$ in Fig. \ref{fig:potentials} also has the same $\omega_{\rm v}$.

The electric dipole function along the vibrational coordinate is modeled as \cite{Hernandez2019,Triana2020}
\begin{equation}
\hat{d}(q)=d_{0}(q-q_{0})e^{-(q-q_{1})^{2}/2\sigma^{2}},
\label{eq:dipole}
\end{equation}
which accurately captures the dipole nature of diatomic and polyatomic molecules, depending on the choice of parameters $d_{0}$, $q_{0}$, $q_{1}$ and $\sigma$ \cite{Grafton2021}. The set of dipole parameters used in this work are given in Table \ref{tab:morse parameters}. These are chosen to describe the behavior of non-polar vibrations ($d_{\rm NP}$), which due to nuclear symmetry have a vanishing dipole moment at equilibrium ($q_\mathrm{eq}$), but acquire a finite dipole moment away from the equilibrium geometry. Examples of this behavior include the degenerate CO stretching mode in iron pentacarbonyl \cite{George2016}. The other dipole behavior of interest corresponds to {\it polar-right} molecules \cite{Triana2020}, which have a finite dipole moment at equilibrium that further increases as the mode distance $q$ increases. Examples of this behavior include lithium fluoride \cite{Triana2018}. In Fig.  \ref{fig:potentials}b we show the dipole functions that correspond to the parameters in Table \ref{tab:morse parameters}. Below we study combinations of these potentials and dipole functions to compute the unitary light-matter interaction dynamics of frequency-driven cavity fields.

\subsection{MCTDH unitary polariton propagation}

For accurately computing the quantum field statistics of cavity photons subject to ultrafast frequency modulation, we solve the time-dependent Schr\"odinger equation in coordinate space with a Hamiltonian given by Eq. (\ref{eq:hamiltonian}) using the multi-configurational time-dependent Hartree (MCTDH) method \cite{mctdhpaper,mctdhbook}, explicitly developed for the accurate and scalable treatment of strongly coupled anharmonic oscillators, as they commonly occur in photochemistry \cite{Vendrell2007}. The MCTDH method has been successfully used to study strong and ultrastrong light-matter coupling of realistic molecules with quantized optical \cite{Vendrell2018,Triana2018} and infrared cavity fields \cite{Triana2020,Triana2020sd}.

In MCTDH, the  system wave function $\Psi(t)$ is written in continuous coordinate space by the ansatz
\begin{equation}
\Psi(q,x,t) = \sum_{j_{q}=1}^{n_{q}} \sum_{j_{x}=1}^{n_{x}} A_{j_{q}j_{x}}(t)\phi_{j_{q}}(q,t)\phi_{j_{x}}(x,t),
\label{eq:mctdh}
\end{equation}
which is an expansion in products of the molecular basis functions $\phi_{j_{q}}(q,t)$ and cavity mode functions $\phi_{j_{x}}(x,t)$. 
To obtain the results below, we set $n_{q}=n_{x}=12$ and represent the vibrational motion in an uniform grid with $N_{q}=721$ grid points on the interval $2.5<q<20.5$ Bohr. The cavity mode is described by a harmonic oscillator basis in a uniform grid with $N_{x}=361$ grid points along the dimensionless interval $-90<x<90$, which is enough to capture the Fock states that participate in the dynamics.

The wavefunction ansatz in Eq. (\ref{eq:mctdh}) is inserted in the Schr\"odinger equation to obtain a set of coupled nonlinear equations for the amplitudes $A_{j_{q}j_{x}}(t)$ and one-dimensional functions $\phi_{j_q}(q,t)$ and $\phi_{j_x}(x,t)$, which are solved numerically using Runge-Kutta propagators.  Eigenstates of the coupled system Hamiltonian can also be obtained in the MCTDH method via propagation in imaginary time \cite{Meyer2003}, which is used  to initialize the coordinate-space wavefunction $\Psi(q,x,t)$ into a Hilbert space vibrational polariton eigenstate \cite{Meyer2006,Vendrell2007,Triana2020}. 

\begin{figure*}[t]
\centering
\includegraphics[width=1\textwidth]{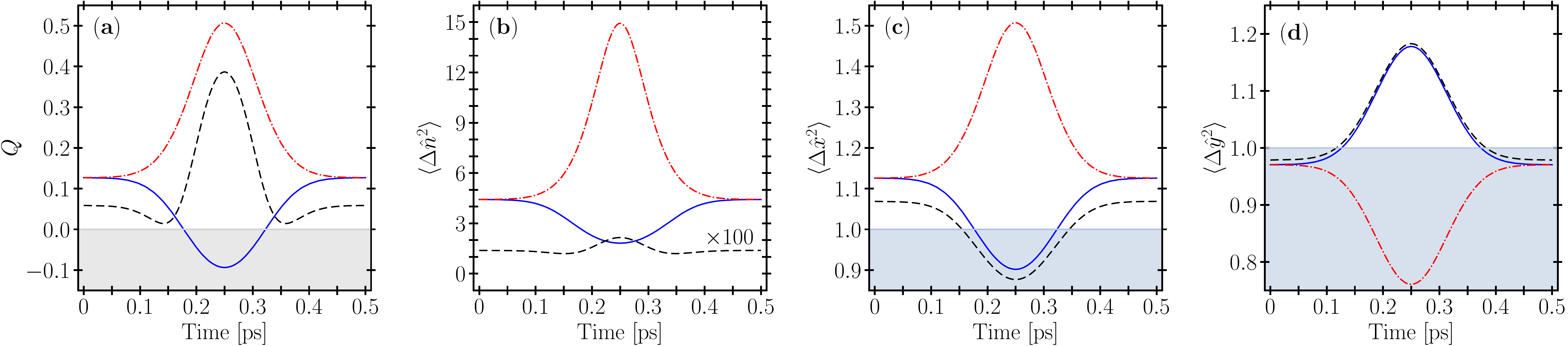}
\caption{
\textbf{Ground state in ultrastrong coupling under frequency modulation.} (a) Evolution of the Mandel $Q$ parameter for a polar-right vibration with a blue modulation of the cavity frequency ($\eta=+0.2$, solid line), a polar-right vibration in red detuning ($\eta=-0.2$, dot-dashed line) and a non-polar vibration in blue detuning ($\eta=+0.2$, dashed line). The shaded region indicates sub-Poissonian photocount statistics. (b) Evolution of the photon number variance $\langle\Delta\hat{n}^{2}\rangle$ for the three cases in panel a. The curve for non-polar vibrations is rescaled by 100. (c) Evolution of the quadrature variance in the phase space direction $\theta=0$. (d) Evolution of the quadrature variance at $\theta=\pi/2$. In panels c and d, the shaded region indicates  squeezing relative to the standard vacuum. The light-matter coupling parameter is $\lambda_g=0.2$. Throughout this work, the frequency modulation pulse is centered at $250$ fs and is 147 fs wide (FWHM).
}
\label{fig:ground}
\end{figure*}

\subsection{Cavity Field Statistics}

The study of intensity and field quadrature correlations is used in quantum optics to distinguish classical and non-classical light sources \cite{Barnett-Radmore}. The normalized intensity correlation at zero delay $g^{(2)}(0)$ is commonly used to identify light sources in which photons are bunched or anti-bunched \cite{Gerry2005}. This photocounting behavior is quantified by the Mandel $Q$ parameter of the intracavity field, given by \cite{Mandel1982} 
\begin{equation}
Q=\frac{\meanval{\Delta\hat{n}^{2}}-\meanval{\hat{n}}}{\meanval{\hat{n}}},
\label{eq:mandel}
\end{equation}
where $\langle\hat{n}\rangle\equiv \langle \hat a^\dagger \hat a\rangle$ is the expectation value of photon number operator and $\langle\Delta\hat{n}^2\rangle$ is the  photon number variance. For a classical laser source that has a Poissonian photocount distribution, the Mandel parameter is $Q=0$ \cite{Gerry2005}.  For sub-Poissonian light sources we have $-1<Q<0$ (Fock state, squeezed vacuum) and for super-Poissonian sources we have $Q>0$ (e.g., thermal light, entangled photon pairs). 

Phase space squeezing is an alternative metric for characterizing quantum light sources. This is done by comparing the variances of the electric field quadratures $\hat x=(\hat a+\hat a^\dagger)$ and $\hat y= -i(\hat a-\hat a^\dagger)$ relative to the vacuum Fock state \cite{Gerry2005}. For vacuum, the normalized quadrature variances are $\langle\Delta\hat{x}^2\rangle=\langle\Delta\hat{y}^2\rangle=1$. Quadrature squeezing implies that $\langle\Delta\hat{x}^2\rangle<1$ or $\langle\Delta\hat{y}^2\rangle<1$, i.e., one of the quadratures components has less quantum noise than the standard vacuum. In terms of the generalized phase space quadrature $\hat X_\theta=\cos\theta\,\hat x+\sin{\theta}\,\hat y$, the squeezing factor (in dB) is defined as $\zeta_\theta=-10\times {\log}\langle\Delta X_\theta^2 \rangle$ \cite{Andersen2016}, where $\theta$ is a phase space angle that can be experimentally tuned using homodyne techniques. We only consider cases with $\theta=0$ and $\theta=\pi/2$ in this article.

\begin{figure*}[t]
\centering
\includegraphics[width=0.89\textwidth]{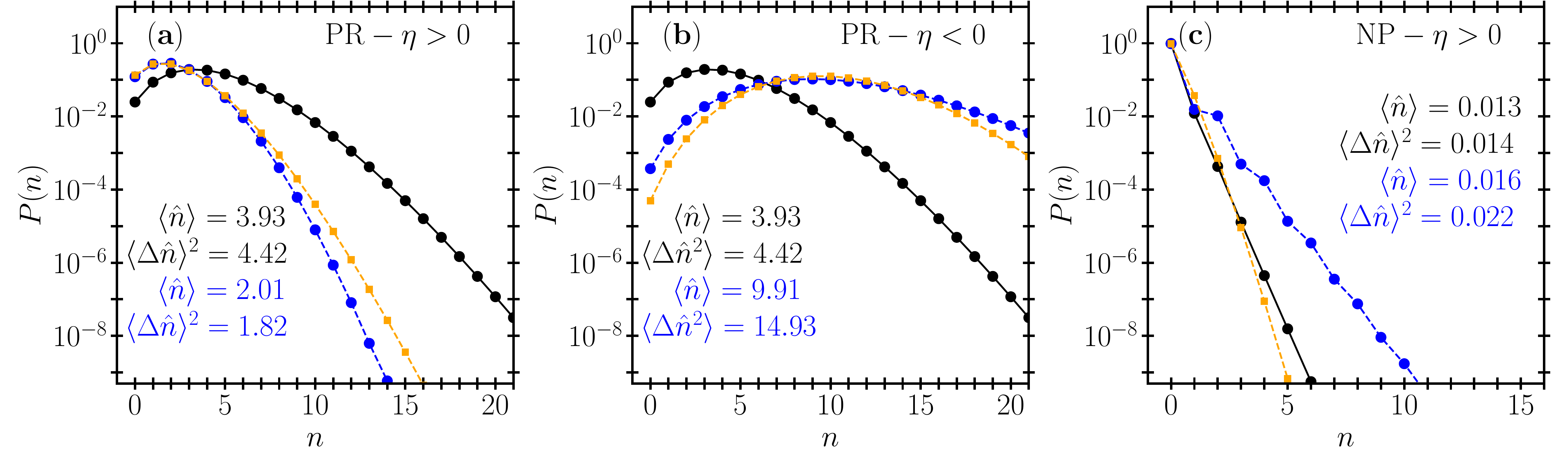}
\caption{
\textbf{Ground state photocount distribution under frequency modulation.} (a) Photon number distribution $P(n)$ before modulation (black line) and the modulation peak (blue line) for a polar-right molecule subject to blue Gaussian frequency modulation ($\eta=+0.2$); (b) $P(n)$ for polar-right molecules with red modulation ($\eta=-0.2$); (c) $P(n)$ for non-polar molecules with blue modulation ($\eta=+0.2$). In all panels, the photon distribution of a coherent state $|\alpha\rangle$ with the same mean photon number as the modulated state at peak detuning is also shown (orange line). The legends indicate the mean photon number and number variance before modulation (black font) and at the modulation peak (blue font). The light-matter coupling parameter is $\lambda_g=0.2$. 
}
\label{fig:groundp}
\end{figure*}

\section{Cavity Frequency Modulation in Ultrastrong Coupling}
\label{sec:vusc}

In this Section we study the evolution of the Mandel $Q$ parameter, the cavity number variance $\langle \Delta \hat n^2\rangle$, and the generalized quadrature variance $\langle \Delta X_\theta^2\rangle $ of the intracavity field $\hat a$, for a coupled vibration-cavity system with a photon frequency modulated according to Eq. (\ref{eq:cavfreq}). The ultrastrong coupling regime is assumed ($\lambda_g>0.1$) for a system initialized in the ground state $\ket{\Psi_{\rm G}}$ or the lower polariton state $\ket{\Psi_{\rm LP}}$ prior to the modulation pulse. We compare the unitary cavity field evolutions obtained for different types of dipole functions and different signs of the peak modulation parameter $\eta$. In Sec. \ref{sec:vsc}, a similar analysis is carried out for a strong coupling scenario ($\lambda_g\ll 0.1$).

\subsection{Initial Ground State}
\label{sec:ground state}

In Figure \ref{fig:ground} we show the evolution of the Mandel $Q$ parameter, number variance, and quadrature variances at $\theta=0$ and $\pi/2$, for a coupled vibration-cavity system that is initially in the ground state $\ket{\Psi_{\rm G}}$ before modulation. The Rabi coupling parameter is $\lambda_g=0.2$ and the cavity is initially on resonance with the fundamental vibration frequency $\omega_{\rm v}$. Three cases are highlighted, differing in the type dipole function $d(q)$ and peak frequency shift $\eta$: polar-right vibration with blue cavity detuning ($\eta>0$);  polar-right vibration with red cavity detuning ($\eta<0$), and  non-polar vibration with blue cavity detuning ($\eta>0$). We use the Morse potential $V_A$ in Table \ref{tab:morse parameters} in all cases. 

Figure \ref{fig:ground} shows that without frequency modulation, the ground state for polar-right vibrations is slightly super-Poissonian ($Q\approx 0.1$), the number variance is relatively large ($\approx 4.5$ photons), there is no squeezing in the $x$-direction, and the $y$-quadrature is only weakly squeezed ($\zeta_{\pi/2}=0.13$ dB) for the chosen $\lambda_g$. For non-polar vibrations, the ground state behaves qualitatively similar to the polar-right case, but the number variance is two orders of magnitude smaller. Significant differences between polar and non-polar species become more evident under frequency modulation.

For polar-right molecules we find an asymmetry in the evolution of the cavity field observables depending on the sign of the modulation amplitude $\eta$. For blue frequency modulations ($\eta>0$), the ground state becomes transiently sub-Poissonian at the modulation peak (see Fig. \ref{fig:ground}a), largely driven by a significant decrease of the number variance (see Fig. \ref{fig:ground}b). The $x$-quadrature also becomes squeezed by $0.41$ dB at the modulation peak (see Fig. \ref{fig:ground}c), accompanied by the increase in the $y$-quadrature variance (see Fig. \ref{fig:ground}d). In contrast, for red modulations ($\eta<0$) the same polar-right vibrational mode generates a cavity field with super-Poissonian statistics ($Q\approx 0.5$) at the modulation peak, associated with a significant increase in the  number variance (see Fig. \ref{fig:ground}b). The $x$-quadrature variance grows relative to the modulation-free value, and the conjugate $y$-quadrature becomes squeezed $\zeta_{\pi/2}=1.2$ dB at the modulation peak.

In order to understand  this behavior, in Fig. \ref{fig:groundp} we show the corresponding photocount distributions $P(n)$ before and during the cavity frequency modulation. In Fig. \ref{fig:groundp}a, we show the decrease of the average photon number and variance predicted for polar-right molecules at the peak of a blue modulation, relative to the initial coupled ground state. This decrease is responsible for the change in the $Q$ parameter around the Poissonian $Q=1$ limit in Fig. \ref{fig:ground}a. For comparison, we also show the exact Poissonian distribution for a coherent state with the same average photon number as the system at the modulation peak. For blue modulations the overlap of the driven system with the coherent state distribution can be significant, but for red modulations the deviations from a coherent state are large at low $n$ (see  Fig. \ref{fig:groundp}b).

The $P(n)$ distributions for non-polar vibrations with and without frequency modulation are shown in Fig. \ref{fig:groundp}c. Without modulation, the photon number and number variance of the coupled ground state are orders of magnitude smaller than the case of polar vibrations. This is consistent with previous works \cite{Hernandez2019,Triana2020}. Despite the small absolute numbers, Fig. \ref{fig:groundp}c shows that the photocount distribution broadens at the peak of a blue modulation shown, which explains the increase of the number variance in Fig. \ref{fig:ground}b.


\subsection{Initial Lower Polariton}
\label{sec:lower polariton}

We now carry out the analysis from the previous section on a coupled system that is initialized in the lower polariton state $\ket{\Psi_{\rm LP}}$ at ultrastrong coupling ($\lambda_g = 0.2$). Excited  polariton population is commonly produced in non-linear cavity infrared spectroscopy experiments \cite{Grafton2021}. We again study polar-right and non-polar vibrations under blue frequency modulation, but now also analyze the dynamics of non-polar molecules under red modulation $(\eta=-0.2)$.

\begin{figure*}[t]
\centering
\includegraphics[width=1\textwidth]{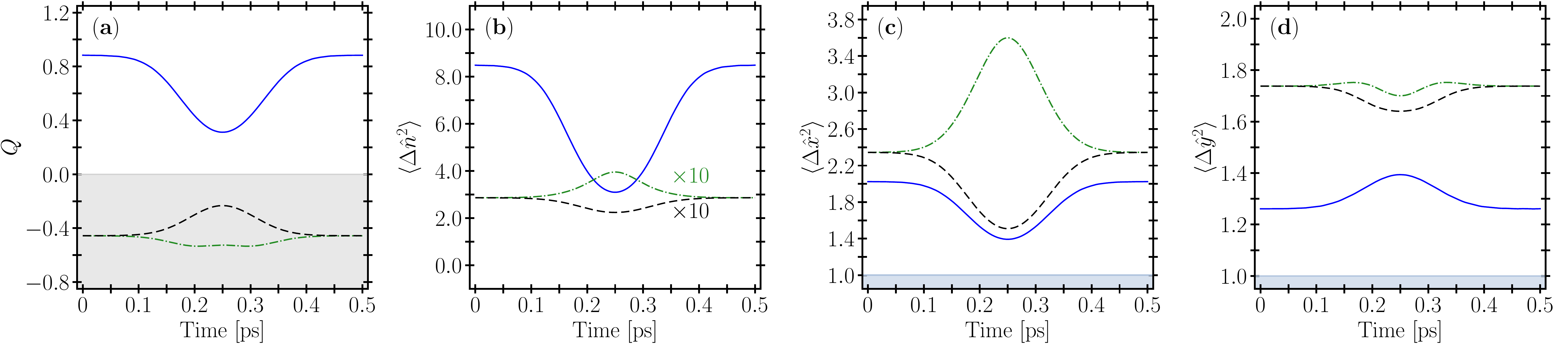}
\caption{
\textbf{Lower polariton in ultrastrong coupling under  frequency modulation.} (a) Evolution of the Mandel $Q$ parameter for a polar-right vibration with a blue modulation of the cavity frequency ($\eta=+0.2$, solid line), a non-polar vibration with red modulation ($\eta=-0.2$, dot-dashed line) and a non-polar vibration with blue modulation ($\eta=+0.2$, dashed line). The shaded region indicates sub-Poissonian photocount statistics. (b) Evolution of the photon number variance $\langle\Delta\hat{n}^{2}\rangle$ for the three cases in panel a. The curve for non-polar vibrations is rescaled by 10. (c) Evolution of the quadrature variance in the phase space direction $\theta=0$. (d) Evolution of the quadrature variance at $\theta=\pi/2$. In panels c and d, the shaded region indicates  squeezing relative to the standard vacuum. The light-matter coupling parameter is $\lambda_g=0.2$. 
}
\label{fig:lower}
\end{figure*}

\begin{figure*}[t]
\centering
\includegraphics[width=0.9\textwidth]{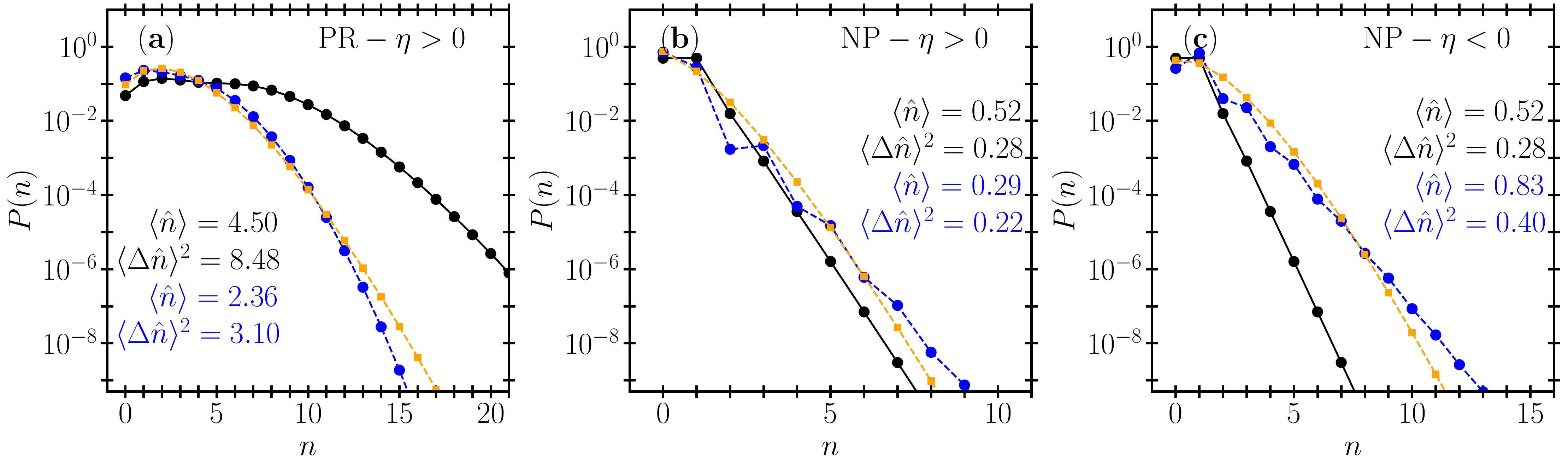}
\caption{
\textbf{Lower polariton photocount distribution under frequency modulation.} (a) Photon number distribution $P(n)$ before modulation (black line) and the modulation peak (blue line) for a polar-right molecule subject to blue Gaussian frequency modulation ($\eta=+0.2$); (b) $P(n)$ for non-polar molecules with blue modulation ($\eta=+0.2$); (c) $P(n)$ for non-polar molecules with red modulation ($\eta=-0.2$). In all panels, the photon distribution of a coherent state $|\alpha\rangle$ with the same mean photon number as the modulated state at peak detuning is also shown (orange line). The legends indicate the mean photon number and number variance before modulation (black font) and at the modulation peak (blue font). The light-matter coupling parameter is $\lambda_g=0.2$. 
}
\label{fig:lowerp}
\end{figure*}

Figure \ref{fig:lower}a shows that lower polariton state for $\lambda_g=0.2$ is super-Poissonian ($Q\approx 0.85$) for polar-right molecules and sub-Poissonian for non-polar vibrations ($Q\approx -0.45$). The number variance is large for polar-right vibrations ($\approx 8.5$ photons), but is an order of magnitude smaller for non-polar species (see Fig. \ref{fig:lower}b). Moreover, there is no  squeezing in the $x$- or $y$- quadrature in either case. Under pulsed frequency modulation, the cavity observables vary significantly for the molecular species considered, although the sign of the Mandel $Q$ parameter does not change for the pulse parameters used. In contrast with the ground state analysis  in Sec. \ref{sec:ground state}, frequency modulating the lower polariton does not generate squeezing at the modulation peak (see Figs. \ref{fig:lower}c and \ref{fig:lower}d). The corresponding photocount distributions with and without modulation are shown in Fig. \ref{fig:lowerp}. The deviation of the number distribution from a classical coherent state is more notorious for both polar and non-polar species.  For non-polar vibrations, Figs. \ref{fig:lowerp}b and \ref{fig:lowerp}c show that the frequency modulation mainly changes the contribution of low-$n$ Fock states.

\section{Non-Adiabatic Excitations in Vibrational Strong Coupling}
\label{sec:vsc}

In Sec. \ref{sec:vusc} we studied the ultrastrong coupling regime of Eq. (\ref{eq:hamiltonian}) by setting $\lambda_g=0.2$. For the Morse potential $V_\mathrm{A}$ (see Table \ref{tab:morse parameters}), this corresponds to the Rabi coupling strength $g_{10}=\lambda_g \omega_{10}\approx 360$ cm$^{-1}$ and a Rabi splitting of  $ 973$ cm$^{-1}$ in linear transmission for polar-right vibrations (greater than $2g_{10}$). Although such splitting magnitudes are currently beyond experimental reach with Fabry-Perot resonators \cite{George2016}, expected improvements in mid-infrared resonator technology may facilitate ultrastrong coupling studies in the near future \cite{Askenazi2017}. 

In this Section we explore the cavity field statistics subject to a photon frequency modulation under strong coupling conditions, which are experimentally accessible with current resonator technology \cite{George2015,Long2015}. For this we set $\lambda_g\sim 10^{-2}$, corresponding to Rabi splittings no greater than $300$ cm$^{-1}$. The main qualitative difference expected in strong coupling is that the polariton spectrum can have energy gaps $\Delta E$ that are comparable with the bandwidth of modulation pulse $\Delta\omega=2\sqrt{2\ln 2}/\tau$. For the modulation pulse assumed in this work ($\tau=62.5$ fs) the bandwidth is $\Delta\omega=226.6 $ cm$^{-1}$. Whenever an eigenstate state that has a nearby energy gap comparable or smaller than $\Delta \omega$ is driven in frequency, non-adiabatic excitations can be expected to remain in the system after the pulse is over \cite{Sarandy2004}.

\begin{figure}[!t]
\centering
\includegraphics[width=0.36\textwidth]{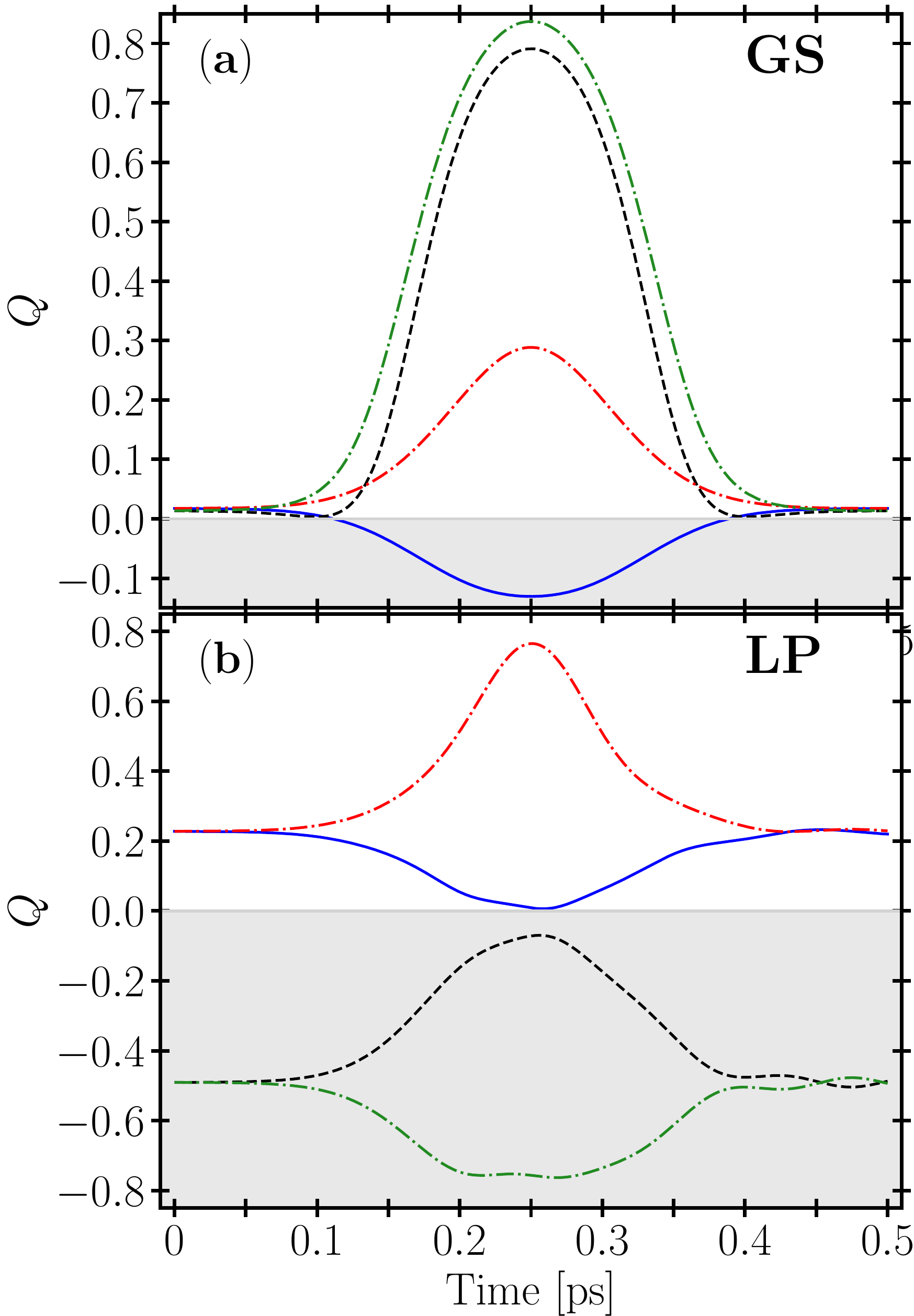}
\caption{
\textbf{Cavity field statistics in vibrational strong coupling}. (a) Evolution of the Mandel $Q$ parameter for a vibration-cavity system initialized in the ground state. (b) Coupled system initialized in the lower polariton state. In all panels, curves correspond to a polar-right vibration with blue modulation ($\eta=+0.2$, solid blue line) and red modulation ($\eta=-0.2$, dot-dashed red line) of the cavity frequency, and for a non-polar vibration with blue detuning ($\eta=+0.2$, dashed black line) and red detuning ($\eta=-0.2$, dot-dashed green line) respect to the undriven cavity frequency. The shaded region indicates sub-Poissonian statistics. The light-matter coupling parameter is $\lambda_g=0.08$.
}
\label{fig:nonadiabatic}
\end{figure}


We illustrate this effect in Figure \ref{fig:nonadiabatic}, where we compare the evolution of the Mandel $Q$ parameter a   vibration-cavity system initialized either in the ground state (Fig. \ref{fig:nonadiabatic}a) or the lower polariton eigenstate  (Fig. \ref{fig:nonadiabatic}b) before the modulation pulse. The coupling parameter is $\lambda_g=0.08$, which would give a Rabi splitting of 288 cm$^{-1}$ in linear transmission. Curves are shown for polar-right and non-polar vibrations under red and blue modulations ($\eta=\pm0.2$). 

Figure \ref{fig:nonadiabatic}a shows that an initial ground state evolves in a qualitatively similar way to the ultrastrong coupling case in Fig. \ref{fig:ground}a, with small differences in the values of $Q$ before and during the modulation pulse. The driven wavefunction returns adiabatically to initial ground state after the pulse is over ($t\approx 0.5$ ps), because the energy gap $\Delta E$ to the first excited (lower) polariton level (1648 cm$^{-1}$ for polar-right vibrations) far exceeds the modulation bandwitdh $\Delta \omega$. In other words, no residual post-pulse excitations remain in the system.

In contrast, Figure \ref{fig:nonadiabatic}b shows that although starting from the lower polariton state can lead to an evolution of the $Q$ parameter with the same trends in Fig. \ref{fig:lower}a for ultrastrong coupling, the pulse turn-off dynamics is qualitatively different. For $\lambda_g=0.08$, the energy gap $\Delta E$ between the lower and upper polariton eigenstates is $307$ cm$^{-1}$ for polar vibrations and $295$ cm$^{-1}$ for non-polar molecules, The latter is  not much greater than the modulation bandwidth $\Delta \omega$. We therefore expect the system wavefunction to exhibit non-adiabatic excitations after the modulation pulse is over. This manifests in Fig. \ref{fig:nonadiabatic}b in an  oscillatory behavior of the Mandel $Q$ parameter as the pulse turns off. 

\begin{figure}[t]
\includegraphics[width=0.36\textwidth]{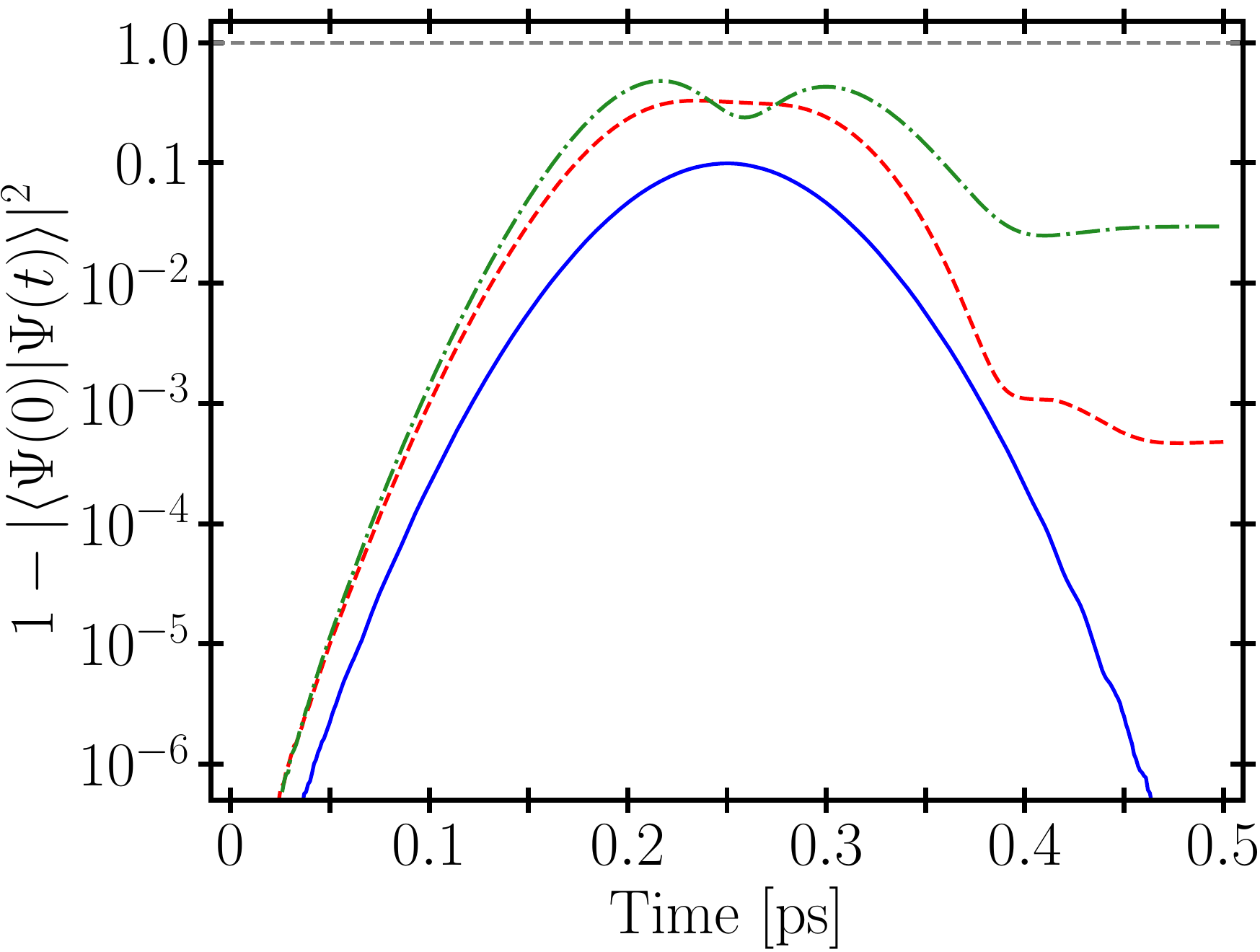}
\caption{\textbf{Non-adiabatic dynamics in vibrational strong coupling.} State autocorrelation for a polar-right vibration subject to cavity frequency modulation for a system initialized in the ground state (solid line), the lower polariton for $\lambda_{g}=0.08$ (dashed line) and the lower polariton with $\lambda_{g}=0.05$ (dot-dashed line). }
\label{fig:autocorrelation}
\end{figure}

We quantify the non-adiabatic excitations generated on the initial lower-polariton wavefunction in Fig. \ref{fig:autocorrelation} for polar-right vibrations under red frequency modulation. Analogue results are found for other dipole functions and values of $\eta$. The degree of adiabaticity of the system evolution is quantified by the autocorrelation function $|\langle \Psi(t)|\Psi(0)\rangle|^2$. For an adiabatic pulse modulation, the post-pulse autocorrelation should not differ from unity. Figure \ref{fig:autocorrelation} shows that this is indeed the case for a driven  ground state (solid line), which as mentioned above is associated with a large energy gap to the first excited polariton level ($\Delta E/\Delta\omega\approx 7.3$). In contrast, for a driven lower polariton state the post-pulse autocorrelation ($t\geq 0.5$) can be significantly different from unity, depending on the magnitude of the Rabi splitting. For example, while for $\lambda_g=0.08$ the system wavefunction is left with about 0.1\% of non-adiabatic excitations after the pulse is over (see Fig. \ref{fig:autocorrelation}, dashed line), for $\lambda_g=0.05$ the residual population outside the lower polariton after the pulse is over is already a few percent (Fig. \ref{fig:autocorrelation}, dot-dashed line), which is feasible to detect in nonlinear infrared spectroscopy \cite{Grafton2021}. This non-adiabatic excitations occur due to the small energy gap  between the lower and upper polariton levels, relative to the modulation bandwidth ($\Delta E/\Delta\omega\approx0.83$ for polar right vibrations).

\section{Dependence on the vibrational anharmonicity}

Molecular vibrations can in principle have similar electric dipole behavior $d(q)$ but differ in the spectral anharmonicity of the nuclear potential. Here we explore the evolution of the Mandel $Q$ parameter in ultrastrong coupling ($\lambda_g=0.2$), for the same frequency modulation in the previous sections. We compare the Morse potentials  Morse potentials $V_\mathrm{A}$ and $V_\mathrm{B}$ in Table \ref{tab:morse parameters}, which have the same fundamental frequency $\omega_{\rm v}$ but different anharmonicity parameters. The latter is quantified by the anharmonic shift $\Delta_{21}\equiv \omega_{\rm v}-\Delta E_{21}$, where $\Delta E_{21}$ is the energy gap between the $v=1$ and $v=2$ vibrational levels. For $V_\mathrm{A}$ we have $\Delta E_{21}=32$ cm$^{-1}$ and for $V_\mathrm{B}$ we have $\Delta_{21}=22$ cm$^{-1}$. For comparison we also consider the purely harmonic case ($\Delta E_{21}=0$). We focus on a system initialized in the ground state, but similar conclusions are found for excited initial states. 

\begin{figure}[t]
\includegraphics[width=0.36\textwidth]{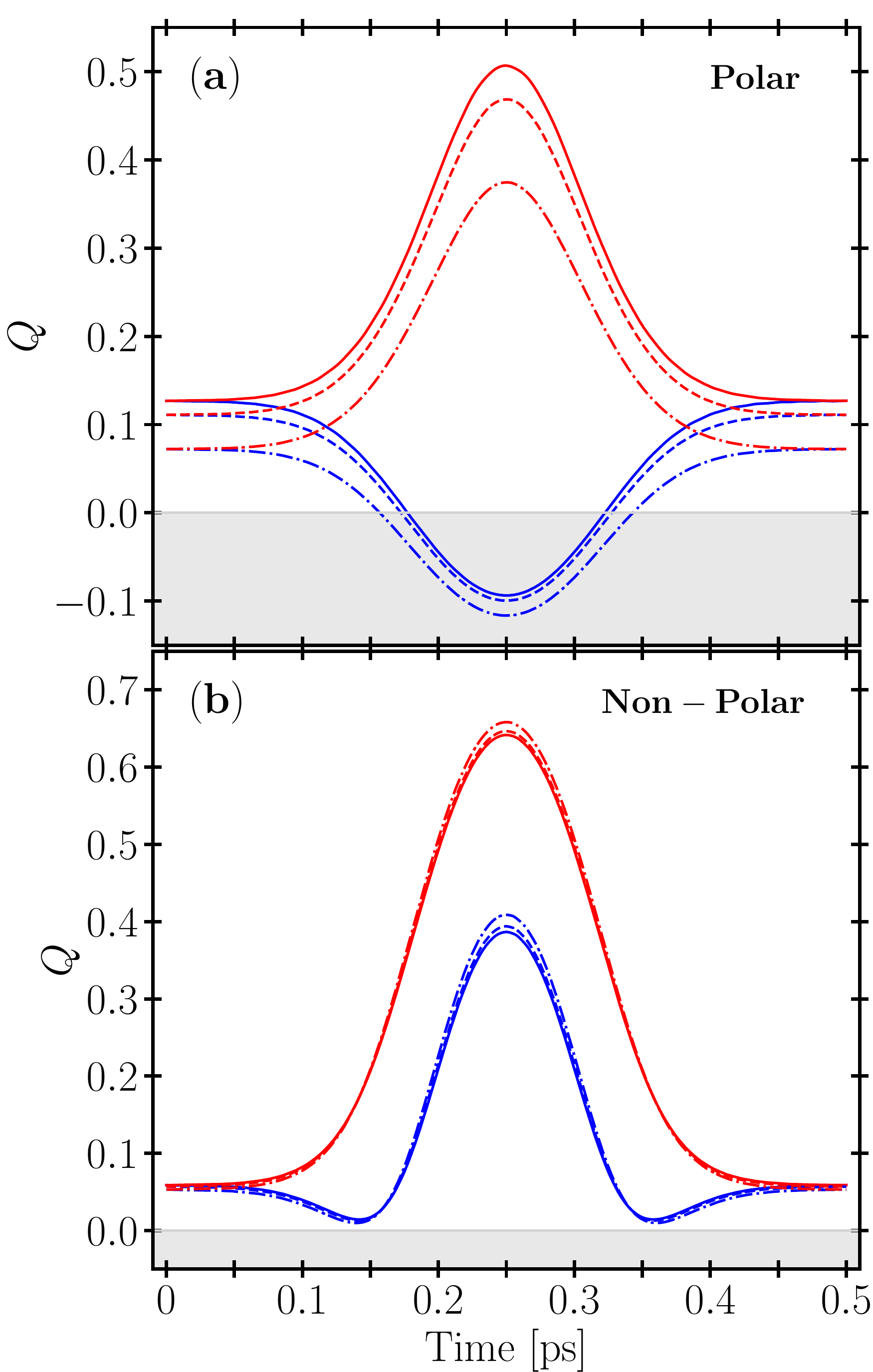}
\caption{\textbf{Cavity field statistics and vibrational anharmonicity.} (a) Evolution of the Mandel $Q$ parameter for a vibration-cavity system initialized in the ground state for a polar-right vibration.  (b) Evolution of the $Q$ parameter for non-polar vibrations. 
In both panels, curves correspond to a molecular system with Morse potentials $V_{\mathrm{A}}$ (solid lines) and $V_{\mathrm{B}}$ (dashed lines), and the harmonic potential $V_{\rm C}$ (dot-dashed lines). 
Red lines correspond to frequency modulations with $\eta=-0.2$ and blue lines to $\eta=+0.2$.
}
\label{fig:harmonic}
\end{figure}

In Fig. \ref{fig:harmonic}a we show the evolution of the $Q$ parameter for polar-right molecules in ultrastrong coupling. The curves for $V_\mathrm{A}$ under blue modulation (blue solid line) and red modulation (red solid line) of the cavity frequency simply reproduce the results in Fig. \ref{fig:ground}a. Reducing the anharmonicity does not qualitatively change the dynamics of the field statistics, i.e., the state can still shift from super-Poissonian to sub-Poissonian for blue modulations ($\eta=0.2$), but the magnitude of the changes are slightly different. For example, the ground state $Q$ parameter before modulation follows the same hierarchy as the degree of anharmonicity, i.e., $Q_\mathrm{A}>Q_\mathrm{B}\geq Q_{\rm HO}$, where $Q_{\rm HO}$ is the harmonic oscillator limit of the ground state at this value of $\lambda_g$. The figure also shows that the relative variations $\Delta Q/Q$ at the modulation peak relative to the initial state, are fairly insensitive to the anharmonicity parameter $\Delta_{21}$. Additional calculations carried out over a broader range values of $\Delta_{21}$ confirm this statement.

Figure \ref{fig:harmonic}b finally shows that at least for a frequency-modulated ground state, no significant  dependence with the spectral anharmonicity is expected for the evolution of the $Q$ parameter for non-polar molecular vibrations. These results are consistent with previous works \cite{Hernandez2019,Triana2020}, which show that light-matter  systems with non-polar molecules are similar to two-level systems and harmonic oscillators with vanishing permanent dipole moments.

\section{Conclusions}  
\label{sec:conclu}

\begin{table*}[!t]
 \caption{\label{tab:cases} \textbf{Summary of predicted cavity field statistics variations.}  Values of the Mandel $Q$ parameter and squeezing factor $\zeta_{\theta}$ at $t=0$ and $t=t_{\mathrm{d}}$ for different dipole moment functions $d(q)$, initial states $|\Psi(0)\rangle$ and cavity frequency modulations ($\eta=\pm0.2$), in the VUSC regime with the light-matter coupling $\lambda_g=0.2$. 
PR and NP correspond to a polar-right and non-polar vibrations, respectively. 
GS and LP correspond to ground and lower polariton states, and the positive and negative sign of $\eta$ represent blue and red modulation, respectively.
}
\begin{ruledtabular}
\begin{tabular}{c c c c  c c c  c c c}
Case & \hspace*{3mm}$d(q)$\hspace*{3mm} & $|\Psi(0)\rangle$ & \hspace*{4mm}$\eta$\hspace*{4mm} & $Q(0)$ & $\zeta_{0}(0)$ & $\zeta_{\pi/2}(0)$ &  $Q(t_{\rm d})$ & $\zeta_{0}(t_{\mathrm{d}})$ & $\zeta_{\pi/2}(t_{\mathrm{d}})$  \\
\hline
\it{i} & PR & GS & $+$ & 0.13 & -0.51 & 0.13 & -0.09 & 0.45 & -0.71   \\
\it{ii} & PR & GS & $-$ & 0.13 & -0.51 & 0.13 & 0.51 & -1.78 & 1.18   \\
\it{iii} & PR & LP & $+$ & 0.88 & -3.06& -1.01 & 0.31 & -1.44 & -1.44   \\
\it{iv} & PR & LP & $-$ &  0.88 & -3.06 & -1.01 & 1.99 & -4.96 & -0.40   \\
\it{v} & NP & GS & $+$ & 0.06 & -0.29 & 0.09 & 0.38 &  0.57 & -0.73    \\
\it{vi} & NP & GS & $-$ & 0.06 & -0.29 & 0.09 & 0.64 &  -1.37 &  1.12   \\
\it{vii} & NP & LP & $+$ & -0.46 & -3.70 & -2.40 & -0.23 & -1.79 & -2.15   \\
\it{viii} & NP & LP & $-$ & -0.46 & -3.70 & -2.40 & -0.53 & -5.56 & -2.31
 \end{tabular}
 \end{ruledtabular}
 \end{table*}

We have shown that the quantum statistics of mid-infrared cavity fields can be dynamically controlled with an ultrafast modulation of the cavity frequency over sub-picosecond timescales, by changing the reflectivity of the conducting cavity mirrors using femtosecond UV pulses.  For $20\%$ shifts of the cavity frequency at the modulation peak, we predict the expected changes in the photocount and quadrature field statistics of the intracavity field by computing the sub-picosecond unitary dynamics of the underlying cavity-vibration polariton Hamiltonian using numerically exact techniques (see Sec. \ref{section:Theory}). 

The dynamics of quantum field statistics is described in terms of the shape of electric dipole function, the anharmonicity of the molecular system, light-matter coupling strength, and whether the system is initially prepared in the ground or lower polariton state before the frequency modulation pulse. For a coupled system in the vibrational ultrastrong coupling (VUSC) regime the cavity field dynamics is adiabatic, while under vibrational strong coupling (VSC) non-adiabatic excitations are possible, provided that the energy gap between the lower and upper polariton levels is comparable with the bandwidth of the frequency modulation pulse.

The results obtained in this work are summarized in Table \ref{tab:cases}, where predicted values of the Mandel $Q$ parameter and squeezing factor $\zeta_{\theta}$ before the modulation ($t=0$) and at the modulation peak ($t=t_\mathrm{d}$) are given for a single Morse potential ($V_\mathrm{A}$) in ultrastrong coupling ($\lambda_g=0.2$), for different polarities of the vibrational mode, different initial states, and different cavity frequency modulations. 

As a general rule, we find that the dipolar nature of molecular vibrations is more important descriptor of the degree of control that can be expected for the quantum field statistics in mid-infrared cavities than the anharmonicity of vibrational spectrum. We also find that the photon statistics for polar-left vibrations \cite{Triana2020} follows the same qualitative behavior with smaller values of $Q$ and $\zeta_{\theta}$ than in the case of polar-right species.
Similarly, if the upper polariton (UP) is considered as the initial state before modulation, the Mandel parameter and squeezing factor behaves similar to the LP case, but the dynamics is reversed, i.e., evolution of the photon statistics under  blue modulation for LP corresponds to dynamics with red modulation for the UP and viceversa.

Cavity designs for implementing the proposed scheme would build on Ref. \cite{Ji2018}, where the reflectivity of the conducting cavity mirrors is modulated by promoting interband transitions with ultrafast UV pulses. This mechanism was recently used to shift the infrared resonance frequency of a semiconducting/insulator  nanoresonator structure \cite{Dunkelberger2020tuning}. Similar design principles could be used with infrared Fabry-Perot cavities with conducting mirror layers \cite{Simpkins2021}, including protecting layers that prevent UV photons from promoting intracavity molecules to electronic excited states. Given the feasibility of the proposed scheme, this work paves the way for the development of integrated molecule-assisted quantum light sources for applications in quantum metrology and photonic quantum information processing in the mid-infrared frequency range.

\begin{acknowledgments}
We thank B. Simpkins, A. Dunkelberger and J. Owrutsky for technical discussions. J.F.T.  is supported by ANID Postdoctoral Fellowship No. 3200565.   F.H. is supported by ANID -- FONDECYT Regular No. 1181743 and ANID –  Millennium Science Initiative Program ICN17\_012.
\end{acknowledgments}

\bibliography{noncla}

\begin{thebibliography}{61}%
\makeatletter
\providecommand \@ifxundefined [1]{%
 \@ifx{#1\undefined}
}%
\providecommand \@ifnum [1]{%
 \ifnum #1\expandafter \@firstoftwo
 \else \expandafter \@secondoftwo
 \fi
}%
\providecommand \@ifx [1]{%
 \ifx #1\expandafter \@firstoftwo
 \else \expandafter \@secondoftwo
 \fi
}%
\providecommand \natexlab [1]{#1}%
\providecommand \enquote  [1]{``#1''}%
\providecommand \bibnamefont  [1]{#1}%
\providecommand \bibfnamefont [1]{#1}%
\providecommand \citenamefont [1]{#1}%
\providecommand \href@noop [0]{\@secondoftwo}%
\providecommand \href [0]{\begingroup \@sanitize@url \@href}%
\providecommand \@href[1]{\@@startlink{#1}\@@href}%
\providecommand \@@href[1]{\endgroup#1\@@endlink}%
\providecommand \@sanitize@url [0]{\catcode `\\12\catcode `\$12\catcode
  `\&12\catcode `\#12\catcode `\^12\catcode `\_12\catcode `\%12\relax}%
\providecommand \@@startlink[1]{}%
\providecommand \@@endlink[0]{}%
\providecommand \url  [0]{\begingroup\@sanitize@url \@url }%
\providecommand \@url [1]{\endgroup\@href {#1}{\urlprefix }}%
\providecommand \urlprefix  [0]{URL }%
\providecommand \Eprint [0]{\href }%
\providecommand \doibase [0]{https://doi.org/}%
\providecommand \selectlanguage [0]{\@gobble}%
\providecommand \bibinfo  [0]{\@secondoftwo}%
\providecommand \bibfield  [0]{\@secondoftwo}%
\providecommand \translation [1]{[#1]}%
\providecommand \BibitemOpen [0]{}%
\providecommand \bibitemStop [0]{}%
\providecommand \bibitemNoStop [0]{.\EOS\space}%
\providecommand \EOS [0]{\spacefactor3000\relax}%
\providecommand \BibitemShut  [1]{\csname bibitem#1\endcsname}%
\let\auto@bib@innerbib\@empty
\bibitem [{\citenamefont {O'Brien}\ \emph {et~al.}(2009)\citenamefont
  {O'Brien}, \citenamefont {Furusawa},\ and\ \citenamefont
  {Vu\v{c}kovi\'{c}}}]{Obrien2009}%
  \BibitemOpen
  \bibfield  {author} {\bibinfo {author} {\bibfnamefont {J.~L.}\ \bibnamefont
  {O'Brien}}, \bibinfo {author} {\bibfnamefont {A.}~\bibnamefont {Furusawa}},\
  and\ \bibinfo {author} {\bibfnamefont {J.}~\bibnamefont {Vu\v{c}kovi\'{c}}},\
  }\bibfield  {title} {\bibinfo {title} {Photonic quantum technologies},\
  }\href {https://doi.org/10.1038/nphoton.2009.229} {\bibfield  {journal}
  {\bibinfo  {journal} {Nature Photonics}\ }\textbf {\bibinfo {volume} {3}},\
  \bibinfo {pages} {687} (\bibinfo {year} {2009})}\BibitemShut {NoStop}%
\bibitem [{\citenamefont {Daiss}\ \emph {et~al.}(2019)\citenamefont {Daiss},
  \citenamefont {Welte}, \citenamefont {Hacker}, \citenamefont {Li},\ and\
  \citenamefont {Rempe}}]{Daiss2019}%
  \BibitemOpen
  \bibfield  {author} {\bibinfo {author} {\bibfnamefont {S.}~\bibnamefont
  {Daiss}}, \bibinfo {author} {\bibfnamefont {S.}~\bibnamefont {Welte}},
  \bibinfo {author} {\bibfnamefont {B.}~\bibnamefont {Hacker}}, \bibinfo
  {author} {\bibfnamefont {L.}~\bibnamefont {Li}},\ and\ \bibinfo {author}
  {\bibfnamefont {G.}~\bibnamefont {Rempe}},\ }\bibfield  {title} {\bibinfo
  {title} {Single-photon distillation via a photonic parity measurement using
  cavity qed},\ }\href {https://doi.org/10.1103/PhysRevLett.122.133603}
  {\bibfield  {journal} {\bibinfo  {journal} {Phys. Rev. Lett.}\ }\textbf
  {\bibinfo {volume} {122}},\ \bibinfo {pages} {133603} (\bibinfo {year}
  {2019})}\BibitemShut {NoStop}%
\bibitem [{\citenamefont {Weiher}\ \emph {et~al.}(2019)\citenamefont {Weiher},
  \citenamefont {Agudelo},\ and\ \citenamefont {Bohmann}}]{Agudelo2019}%
  \BibitemOpen
  \bibfield  {author} {\bibinfo {author} {\bibfnamefont {K.}~\bibnamefont
  {Weiher}}, \bibinfo {author} {\bibfnamefont {E.}~\bibnamefont {Agudelo}},\
  and\ \bibinfo {author} {\bibfnamefont {M.}~\bibnamefont {Bohmann}},\
  }\bibfield  {title} {\bibinfo {title} {Conditional nonclassical field
  generation in cavity qed},\ }\href
  {https://doi.org/10.1103/PhysRevA.100.043812} {\bibfield  {journal} {\bibinfo
   {journal} {Phys. Rev. A}\ }\textbf {\bibinfo {volume} {100}},\ \bibinfo
  {pages} {043812} (\bibinfo {year} {2019})}\BibitemShut {NoStop}%
\bibitem [{\citenamefont {Aspelmeyer}\ \emph {et~al.}(2014)\citenamefont
  {Aspelmeyer}, \citenamefont {Kippenberg},\ and\ \citenamefont
  {Marquardt}}]{Aspelmeyer2014}%
  \BibitemOpen
  \bibfield  {author} {\bibinfo {author} {\bibfnamefont {M.}~\bibnamefont
  {Aspelmeyer}}, \bibinfo {author} {\bibfnamefont {T.~J.}\ \bibnamefont
  {Kippenberg}},\ and\ \bibinfo {author} {\bibfnamefont {F.}~\bibnamefont
  {Marquardt}},\ }\bibfield  {title} {\bibinfo {title} {Cavity optomechanics},\
  }\href {https://doi.org/10.1103/RevModPhys.86.1391} {\bibfield  {journal}
  {\bibinfo  {journal} {Rev. Mod. Phys.}\ }\textbf {\bibinfo {volume} {86}},\
  \bibinfo {pages} {1391} (\bibinfo {year} {2014})}\BibitemShut {NoStop}%
\bibitem [{\citenamefont {S\'aez-Bl\'azquez}\ \emph {et~al.}(2018)\citenamefont
  {S\'aez-Bl\'azquez}, \citenamefont {Feist}, \citenamefont
  {Garc\'{\i}a-Vidal},\ and\ \citenamefont
  {Fern\'andez-Dom\'{\i}nguez}}]{Saez2018}%
  \BibitemOpen
  \bibfield  {author} {\bibinfo {author} {\bibfnamefont {R.}~\bibnamefont
  {S\'aez-Bl\'azquez}}, \bibinfo {author} {\bibfnamefont {J.}~\bibnamefont
  {Feist}}, \bibinfo {author} {\bibfnamefont {F.~J.}\ \bibnamefont
  {Garc\'{\i}a-Vidal}},\ and\ \bibinfo {author} {\bibfnamefont {A.~I.}\
  \bibnamefont {Fern\'andez-Dom\'{\i}nguez}},\ }\bibfield  {title} {\bibinfo
  {title} {Photon statistics in collective strong coupling: Nanocavities and
  microcavities},\ }\href {https://doi.org/10.1103/PhysRevA.98.013839}
  {\bibfield  {journal} {\bibinfo  {journal} {Phys. Rev. A}\ }\textbf {\bibinfo
  {volume} {98}},\ \bibinfo {pages} {013839} (\bibinfo {year}
  {2018})}\BibitemShut {NoStop}%
\bibitem [{\citenamefont {Kimble}(2008)}]{Kimble2008}%
  \BibitemOpen
  \bibfield  {author} {\bibinfo {author} {\bibfnamefont {H.~J.}\ \bibnamefont
  {Kimble}},\ }\bibfield  {title} {\bibinfo {title} {The quantum internet},\
  }\href {https://doi.org/10.1038/nature07127} {\bibfield  {journal} {\bibinfo
  {journal} {Nature}\ }\textbf {\bibinfo {volume} {453}},\ \bibinfo {pages}
  {1023} (\bibinfo {year} {2008})}\BibitemShut {NoStop}%
\bibitem [{\citenamefont {Yu}\ \emph {et~al.}(2015)\citenamefont {Yu},
  \citenamefont {Natarajan}, \citenamefont {Horikiri}, \citenamefont
  {Langrock}, \citenamefont {Pelc}, \citenamefont {Tanner}, \citenamefont
  {Abe}, \citenamefont {Maier}, \citenamefont {Schneider}, \citenamefont
  {H{\"o}fling}, \citenamefont {Kamp}, \citenamefont {Hadfield}, \citenamefont
  {Fejer},\ and\ \citenamefont {Yamamoto}}]{Yu2015}%
  \BibitemOpen
  \bibfield  {author} {\bibinfo {author} {\bibfnamefont {L.}~\bibnamefont
  {Yu}}, \bibinfo {author} {\bibfnamefont {C.~M.}\ \bibnamefont {Natarajan}},
  \bibinfo {author} {\bibfnamefont {T.}~\bibnamefont {Horikiri}}, \bibinfo
  {author} {\bibfnamefont {C.}~\bibnamefont {Langrock}}, \bibinfo {author}
  {\bibfnamefont {J.~S.}\ \bibnamefont {Pelc}}, \bibinfo {author}
  {\bibfnamefont {M.~G.}\ \bibnamefont {Tanner}}, \bibinfo {author}
  {\bibfnamefont {E.}~\bibnamefont {Abe}}, \bibinfo {author} {\bibfnamefont
  {S.}~\bibnamefont {Maier}}, \bibinfo {author} {\bibfnamefont
  {C.}~\bibnamefont {Schneider}}, \bibinfo {author} {\bibfnamefont
  {S.}~\bibnamefont {H{\"o}fling}}, \bibinfo {author} {\bibfnamefont
  {M.}~\bibnamefont {Kamp}}, \bibinfo {author} {\bibfnamefont {R.~H.}\
  \bibnamefont {Hadfield}}, \bibinfo {author} {\bibfnamefont {M.~M.}\
  \bibnamefont {Fejer}},\ and\ \bibinfo {author} {\bibfnamefont
  {Y.}~\bibnamefont {Yamamoto}},\ }\bibfield  {title} {\bibinfo {title}
  {Two-photon interference at telecom wavelengths for time-bin-encoded single
  photons from quantum-dot spin qubits},\ }\href
  {https://doi.org/10.1038/ncomms9955} {\bibfield  {journal} {\bibinfo
  {journal} {Nature Communications}\ }\textbf {\bibinfo {volume} {6}},\
  \bibinfo {pages} {8955} (\bibinfo {year} {2015})}\BibitemShut {NoStop}%
\bibitem [{\citenamefont {Clark}\ \emph {et~al.}(2021)\citenamefont {Clark},
  \citenamefont {Chekhova}, \citenamefont {Matthews}, \citenamefont {Rarity},\
  and\ \citenamefont {Oulton}}]{Clark2021}%
  \BibitemOpen
  \bibfield  {author} {\bibinfo {author} {\bibfnamefont {A.~S.}\ \bibnamefont
  {Clark}}, \bibinfo {author} {\bibfnamefont {M.}~\bibnamefont {Chekhova}},
  \bibinfo {author} {\bibfnamefont {J.~C.~F.}\ \bibnamefont {Matthews}},
  \bibinfo {author} {\bibfnamefont {J.~G.}\ \bibnamefont {Rarity}},\ and\
  \bibinfo {author} {\bibfnamefont {R.~F.}\ \bibnamefont {Oulton}},\ }\bibfield
   {title} {\bibinfo {title} {Special topic: Quantum sensing with correlated
  light sources},\ }\href {https://doi.org/10.1063/5.0041043} {\bibfield
  {journal} {\bibinfo  {journal} {Applied Physics Letters}\ }\textbf {\bibinfo
  {volume} {118}},\ \bibinfo {pages} {060401} (\bibinfo {year} {2021})},\
  \Eprint {https://arxiv.org/abs/https://doi.org/10.1063/5.0041043}
  {https://doi.org/10.1063/5.0041043} \BibitemShut {NoStop}%
\bibitem [{\citenamefont {Vega}\ \emph {et~al.}(2020)\citenamefont {Vega},
  \citenamefont {Saravi}, \citenamefont {Pertsch},\ and\ \citenamefont
  {Setzpfandt}}]{Vega2020}%
  \BibitemOpen
  \bibfield  {author} {\bibinfo {author} {\bibfnamefont {A.}~\bibnamefont
  {Vega}}, \bibinfo {author} {\bibfnamefont {S.}~\bibnamefont {Saravi}},
  \bibinfo {author} {\bibfnamefont {T.}~\bibnamefont {Pertsch}},\ and\ \bibinfo
  {author} {\bibfnamefont {F.}~\bibnamefont {Setzpfandt}},\ }\bibfield  {title}
  {\bibinfo {title} {Pinhole quantum ghost imaging},\ }\href
  {https://doi.org/10.1063/5.0012477} {\bibfield  {journal} {\bibinfo
  {journal} {Applied Physics Letters}\ }\textbf {\bibinfo {volume} {117}},\
  \bibinfo {pages} {094003} (\bibinfo {year} {2020})},\ \Eprint
  {https://arxiv.org/abs/https://doi.org/10.1063/5.0012477}
  {https://doi.org/10.1063/5.0012477} \BibitemShut {NoStop}%
\bibitem [{\citenamefont {Zhang}\ \emph {et~al.}(2015)\citenamefont {Zhang},
  \citenamefont {Itzler}, \citenamefont {Zbinden},\ and\ \citenamefont
  {Pan}}]{Zhang2015}%
  \BibitemOpen
  \bibfield  {author} {\bibinfo {author} {\bibfnamefont {J.}~\bibnamefont
  {Zhang}}, \bibinfo {author} {\bibfnamefont {M.~A.}\ \bibnamefont {Itzler}},
  \bibinfo {author} {\bibfnamefont {H.}~\bibnamefont {Zbinden}},\ and\ \bibinfo
  {author} {\bibfnamefont {J.-W.}\ \bibnamefont {Pan}},\ }\bibfield  {title}
  {\bibinfo {title} {Advances in ingaas/inp single-photon detector systems for
  quantum communication},\ }\href {https://doi.org/10.1038/lsa.2015.59}
  {\bibfield  {journal} {\bibinfo  {journal} {Light: Science \& Applications}\
  }\textbf {\bibinfo {volume} {4}},\ \bibinfo {pages} {e286} (\bibinfo {year}
  {2015})}\BibitemShut {NoStop}%
\bibitem [{\citenamefont {Natarajan}\ \emph {et~al.}(2012)\citenamefont
  {Natarajan}, \citenamefont {Tanner},\ and\ \citenamefont
  {Hadfield}}]{Natarajan2012}%
  \BibitemOpen
  \bibfield  {author} {\bibinfo {author} {\bibfnamefont {C.~M.}\ \bibnamefont
  {Natarajan}}, \bibinfo {author} {\bibfnamefont {M.~G.}\ \bibnamefont
  {Tanner}},\ and\ \bibinfo {author} {\bibfnamefont {R.~H.}\ \bibnamefont
  {Hadfield}},\ }\bibfield  {title} {\bibinfo {title} {Superconducting nanowire
  single-photon detectors: physics and applications},\ }\href
  {https://doi.org/10.1088/0953-2048/25/6/063001} {\bibfield  {journal}
  {\bibinfo  {journal} {Superconductor Science and Technology}\ }\textbf
  {\bibinfo {volume} {25}},\ \bibinfo {pages} {063001} (\bibinfo {year}
  {2012})}\BibitemShut {NoStop}%
\bibitem [{\citenamefont {Shields}\ \emph {et~al.}(2020)\citenamefont
  {Shields}, \citenamefont {Prabhakar}, \citenamefont {Dada}, \citenamefont
  {Ebrahim}, \citenamefont {Taylor}, \citenamefont {Morozov}, \citenamefont
  {Erotokritou}, \citenamefont {Miki}, \citenamefont {Yabuno}, \citenamefont
  {Terai}, \citenamefont {Gawith}, \citenamefont {Kues}, \citenamefont
  {Caspani}, \citenamefont {Hadfield},\ and\ \citenamefont
  {Clerici}}]{Shields2020}%
  \BibitemOpen
  \bibfield  {author} {\bibinfo {author} {\bibfnamefont {T.}~\bibnamefont
  {Shields}}, \bibinfo {author} {\bibfnamefont {S.}~\bibnamefont {Prabhakar}},
  \bibinfo {author} {\bibfnamefont {A.}~\bibnamefont {Dada}}, \bibinfo {author}
  {\bibfnamefont {M.}~\bibnamefont {Ebrahim}}, \bibinfo {author} {\bibfnamefont
  {G.~G.}\ \bibnamefont {Taylor}}, \bibinfo {author} {\bibfnamefont
  {D.}~\bibnamefont {Morozov}}, \bibinfo {author} {\bibfnamefont
  {K.}~\bibnamefont {Erotokritou}}, \bibinfo {author} {\bibfnamefont
  {S.}~\bibnamefont {Miki}}, \bibinfo {author} {\bibfnamefont {M.}~\bibnamefont
  {Yabuno}}, \bibinfo {author} {\bibfnamefont {H.}~\bibnamefont {Terai}},
  \bibinfo {author} {\bibfnamefont {C.}~\bibnamefont {Gawith}}, \bibinfo
  {author} {\bibfnamefont {M.}~\bibnamefont {Kues}}, \bibinfo {author}
  {\bibfnamefont {L.}~\bibnamefont {Caspani}}, \bibinfo {author} {\bibfnamefont
  {R.~H.}\ \bibnamefont {Hadfield}},\ and\ \bibinfo {author} {\bibfnamefont
  {M.}~\bibnamefont {Clerici}},\ }\bibfield  {title} {\bibinfo {title}
  {Mid-infrared quantum interference and polarization entanglement},\ }in\
  \href {https://doi.org/10.1364/MICS.2020.MF1C.7} {\emph {\bibinfo {booktitle}
  {OSA High-brightness Sources and Light-driven Interactions Congress 2020
  (EUVXRAY, HILAS, MICS)}}}\ (\bibinfo  {publisher} {Optical Society of
  America},\ \bibinfo {year} {2020})\ p.\ \bibinfo {pages} {MF1C.7}\BibitemShut
  {NoStop}%
\bibitem [{\citenamefont {Spitz}\ \emph {et~al.}(2021)\citenamefont {Spitz},
  \citenamefont {Herdt}, \citenamefont {Wu}, \citenamefont {Maisons},
  \citenamefont {Carras}, \citenamefont {Wong}, \citenamefont
  {Els{\"a}{\ss}er},\ and\ \citenamefont {Grillot}}]{Spitz2021}%
  \BibitemOpen
  \bibfield  {author} {\bibinfo {author} {\bibfnamefont {O.}~\bibnamefont
  {Spitz}}, \bibinfo {author} {\bibfnamefont {A.}~\bibnamefont {Herdt}},
  \bibinfo {author} {\bibfnamefont {J.}~\bibnamefont {Wu}}, \bibinfo {author}
  {\bibfnamefont {G.}~\bibnamefont {Maisons}}, \bibinfo {author} {\bibfnamefont
  {M.}~\bibnamefont {Carras}}, \bibinfo {author} {\bibfnamefont {C.-W.}\
  \bibnamefont {Wong}}, \bibinfo {author} {\bibfnamefont {W.}~\bibnamefont
  {Els{\"a}{\ss}er}},\ and\ \bibinfo {author} {\bibfnamefont {F.}~\bibnamefont
  {Grillot}},\ }\bibfield  {title} {\bibinfo {title} {Private communication
  with quantum cascade laser photonic chaos},\ }\href
  {https://doi.org/10.1038/s41467-021-23527-9} {\bibfield  {journal} {\bibinfo
  {journal} {Nature Communications}\ }\textbf {\bibinfo {volume} {12}},\
  \bibinfo {pages} {3327} (\bibinfo {year} {2021})}\BibitemShut {NoStop}%
\bibitem [{\citenamefont {Yao}\ \emph {et~al.}(2012)\citenamefont {Yao},
  \citenamefont {Hoffman},\ and\ \citenamefont {Gmachl}}]{Yao2012}%
  \BibitemOpen
  \bibfield  {author} {\bibinfo {author} {\bibfnamefont {Y.}~\bibnamefont
  {Yao}}, \bibinfo {author} {\bibfnamefont {A.~J.}\ \bibnamefont {Hoffman}},\
  and\ \bibinfo {author} {\bibfnamefont {C.~F.}\ \bibnamefont {Gmachl}},\
  }\bibfield  {title} {\bibinfo {title} {Mid-infrared quantum cascade lasers},\
  }\href {https://doi.org/10.1038/nphoton.2012.143} {\bibfield  {journal}
  {\bibinfo  {journal} {Nature Photonics}\ }\textbf {\bibinfo {volume} {6}},\
  \bibinfo {pages} {432} (\bibinfo {year} {2012})}\BibitemShut {NoStop}%
\bibitem [{\citenamefont {Gabbrielli}\ \emph {et~al.}(2021)\citenamefont
  {Gabbrielli}, \citenamefont {Cappelli}, \citenamefont {Bruno}, \citenamefont
  {Corrias}, \citenamefont {Borri}, \citenamefont {Natale},\ and\ \citenamefont
  {Zavatta}}]{Gabbrielli2021}%
  \BibitemOpen
  \bibfield  {author} {\bibinfo {author} {\bibfnamefont {T.}~\bibnamefont
  {Gabbrielli}}, \bibinfo {author} {\bibfnamefont {F.}~\bibnamefont
  {Cappelli}}, \bibinfo {author} {\bibfnamefont {N.}~\bibnamefont {Bruno}},
  \bibinfo {author} {\bibfnamefont {N.}~\bibnamefont {Corrias}}, \bibinfo
  {author} {\bibfnamefont {S.}~\bibnamefont {Borri}}, \bibinfo {author}
  {\bibfnamefont {P.~D.}\ \bibnamefont {Natale}},\ and\ \bibinfo {author}
  {\bibfnamefont {A.}~\bibnamefont {Zavatta}},\ }\bibfield  {title} {\bibinfo
  {title} {Mid-infrared homodyne balanced detector for quantum light
  characterization},\ }\href {https://doi.org/10.1364/OE.420990} {\bibfield
  {journal} {\bibinfo  {journal} {Opt. Express}\ }\textbf {\bibinfo {volume}
  {29}},\ \bibinfo {pages} {14536} (\bibinfo {year} {2021})}\BibitemShut
  {NoStop}%
\bibitem [{\citenamefont {Mancinelli}\ \emph {et~al.}(2017)\citenamefont
  {Mancinelli}, \citenamefont {Trenti}, \citenamefont {Piccione}, \citenamefont
  {Fontana}, \citenamefont {Dam}, \citenamefont {Tidemand-Lichtenberg},
  \citenamefont {Pedersen},\ and\ \citenamefont {Pavesi}}]{Mancinelli2017}%
  \BibitemOpen
  \bibfield  {author} {\bibinfo {author} {\bibfnamefont {M.}~\bibnamefont
  {Mancinelli}}, \bibinfo {author} {\bibfnamefont {A.}~\bibnamefont {Trenti}},
  \bibinfo {author} {\bibfnamefont {S.}~\bibnamefont {Piccione}}, \bibinfo
  {author} {\bibfnamefont {G.}~\bibnamefont {Fontana}}, \bibinfo {author}
  {\bibfnamefont {J.~S.}\ \bibnamefont {Dam}}, \bibinfo {author} {\bibfnamefont
  {P.}~\bibnamefont {Tidemand-Lichtenberg}}, \bibinfo {author} {\bibfnamefont
  {C.}~\bibnamefont {Pedersen}},\ and\ \bibinfo {author} {\bibfnamefont
  {L.}~\bibnamefont {Pavesi}},\ }\bibfield  {title} {\bibinfo {title}
  {Mid-infrared coincidence measurements on twin photons at room temperature},\
  }\href {https://doi.org/10.1038/ncomms15184} {\bibfield  {journal} {\bibinfo
  {journal} {Nature Communications}\ }\textbf {\bibinfo {volume} {8}},\
  \bibinfo {pages} {15184} (\bibinfo {year} {2017})}\BibitemShut {NoStop}%
\bibitem [{\citenamefont {Shalabney}\ \emph
  {et~al.}(2015{\natexlab{a}})\citenamefont {Shalabney}, \citenamefont
  {George}, \citenamefont {Hutchison}, \citenamefont {Pupillo}, \citenamefont
  {Genet},\ and\ \citenamefont {Ebbesen}}]{Shalabney2015coherent}%
  \BibitemOpen
  \bibfield  {author} {\bibinfo {author} {\bibfnamefont {A.}~\bibnamefont
  {Shalabney}}, \bibinfo {author} {\bibfnamefont {J.}~\bibnamefont {George}},
  \bibinfo {author} {\bibfnamefont {J.}~\bibnamefont {Hutchison}}, \bibinfo
  {author} {\bibfnamefont {G.}~\bibnamefont {Pupillo}}, \bibinfo {author}
  {\bibfnamefont {C.}~\bibnamefont {Genet}},\ and\ \bibinfo {author}
  {\bibfnamefont {T.~W.}\ \bibnamefont {Ebbesen}},\ }\bibfield  {title}
  {\bibinfo {title} {{Coherent coupling of molecular resonators with a
  microcavity mode}},\ }\href {https://doi.org/10.1038/ncomms6981} {\bibfield
  {journal} {\bibinfo  {journal} {Nature Communications}\ }\textbf {\bibinfo
  {volume} {6}},\ \bibinfo {pages} {1} (\bibinfo {year}
  {2015}{\natexlab{a}})}\BibitemShut {NoStop}%
\bibitem [{\citenamefont {Shalabney}\ \emph
  {et~al.}(2015{\natexlab{b}})\citenamefont {Shalabney}, \citenamefont
  {George}, \citenamefont {Hiura}, \citenamefont {Hutchison}, \citenamefont
  {Genet}, \citenamefont {Hellwig},\ and\ \citenamefont
  {Ebbesen}}]{Shalabney2015raman}%
  \BibitemOpen
  \bibfield  {author} {\bibinfo {author} {\bibfnamefont {A.}~\bibnamefont
  {Shalabney}}, \bibinfo {author} {\bibfnamefont {J.}~\bibnamefont {George}},
  \bibinfo {author} {\bibfnamefont {H.}~\bibnamefont {Hiura}}, \bibinfo
  {author} {\bibfnamefont {J.~A.}\ \bibnamefont {Hutchison}}, \bibinfo {author}
  {\bibfnamefont {C.}~\bibnamefont {Genet}}, \bibinfo {author} {\bibfnamefont
  {P.}~\bibnamefont {Hellwig}},\ and\ \bibinfo {author} {\bibfnamefont {T.~W.}\
  \bibnamefont {Ebbesen}},\ }\bibfield  {title} {\bibinfo {title} {{Enhanced
  Raman Scattering from Vibro-Polariton Hybrid States}},\ }\href
  {https://doi.org/10.1002/anie.201502979} {\bibfield  {journal} {\bibinfo
  {journal} {Angewandte Chemie International Edition}\ }\textbf {\bibinfo
  {volume} {54}},\ \bibinfo {pages} {7971} (\bibinfo {year}
  {2015}{\natexlab{b}})}\BibitemShut {NoStop}%
\bibitem [{\citenamefont {George}\ \emph {et~al.}(2015)\citenamefont {George},
  \citenamefont {Shalabney}, \citenamefont {Hutchison}, \citenamefont {Genet},\
  and\ \citenamefont {Ebbesen}}]{George2015}%
  \BibitemOpen
  \bibfield  {author} {\bibinfo {author} {\bibfnamefont {J.}~\bibnamefont
  {George}}, \bibinfo {author} {\bibfnamefont {A.}~\bibnamefont {Shalabney}},
  \bibinfo {author} {\bibfnamefont {J.~A.}\ \bibnamefont {Hutchison}}, \bibinfo
  {author} {\bibfnamefont {C.}~\bibnamefont {Genet}},\ and\ \bibinfo {author}
  {\bibfnamefont {T.~W.}\ \bibnamefont {Ebbesen}},\ }\bibfield  {title}
  {\bibinfo {title} {{Liquid-phase vibrational strong coupling}},\ }\href
  {https://doi.org/10.1021/acs.jpclett.5b00204} {\bibfield  {journal} {\bibinfo
   {journal} {Journal of Physical Chemistry Letters}\ }\textbf {\bibinfo
  {volume} {6}},\ \bibinfo {pages} {1027} (\bibinfo {year} {2015})}\BibitemShut
  {NoStop}%
\bibitem [{\citenamefont {Long}\ and\ \citenamefont
  {Simpkins}(2015)}]{Long2015}%
  \BibitemOpen
  \bibfield  {author} {\bibinfo {author} {\bibfnamefont {J.~P.}\ \bibnamefont
  {Long}}\ and\ \bibinfo {author} {\bibfnamefont {B.~S.}\ \bibnamefont
  {Simpkins}},\ }\bibfield  {title} {\bibinfo {title} {Coherent coupling
  between a molecular vibration and fabry--perot optical cavity to give
  hybridized states in the strong coupling limit},\ }\href
  {https://doi.org/10.1021/ph5003347} {\bibfield  {journal} {\bibinfo
  {journal} {ACS Photonics}\ }\textbf {\bibinfo {volume} {2}},\ \bibinfo
  {pages} {130} (\bibinfo {year} {2015})}\BibitemShut {NoStop}%
\bibitem [{\citenamefont {Grafton}\ \emph
  {et~al.}(2021{\natexlab{a}})\citenamefont {Grafton}, \citenamefont
  {Dunkelberger}, \citenamefont {Simpkins}, \citenamefont {Triana},
  \citenamefont {Hern{\'{a}}ndez}, \citenamefont {Herrera},\ and\ \citenamefont
  {Owrutsky}}]{Grafton2020}%
  \BibitemOpen
  \bibfield  {author} {\bibinfo {author} {\bibfnamefont {A.~B.}\ \bibnamefont
  {Grafton}}, \bibinfo {author} {\bibfnamefont {A.~D.}\ \bibnamefont
  {Dunkelberger}}, \bibinfo {author} {\bibfnamefont {B.~S.}\ \bibnamefont
  {Simpkins}}, \bibinfo {author} {\bibfnamefont {J.~F.}\ \bibnamefont
  {Triana}}, \bibinfo {author} {\bibfnamefont {F.~J.}\ \bibnamefont
  {Hern{\'{a}}ndez}}, \bibinfo {author} {\bibfnamefont {F.}~\bibnamefont
  {Herrera}},\ and\ \bibinfo {author} {\bibfnamefont {J.~C.}\ \bibnamefont
  {Owrutsky}},\ }\bibfield  {title} {\bibinfo {title} {{Excited-state
  vibration-polariton transitions and dynamics in nitroprusside}},\ }\href
  {https://doi.org/10.1038/s41467-020-20535-z} {\bibfield  {journal} {\bibinfo
  {journal} {Nature Communications}\ }\textbf {\bibinfo {volume} {12}},\
  \bibinfo {pages} {214} (\bibinfo {year} {2021}{\natexlab{a}})}\BibitemShut
  {NoStop}%
\bibitem [{\citenamefont {Xiang}\ \emph {et~al.}(2019)\citenamefont {Xiang},
  \citenamefont {Ribeiro}, \citenamefont {Li}, \citenamefont {Dunkelberger},
  \citenamefont {Simpkins}, \citenamefont {Yuen-Zhou},\ and\ \citenamefont
  {Xiong}}]{Xiang2019manipulating}%
  \BibitemOpen
  \bibfield  {author} {\bibinfo {author} {\bibfnamefont {B.}~\bibnamefont
  {Xiang}}, \bibinfo {author} {\bibfnamefont {R.~F.}\ \bibnamefont {Ribeiro}},
  \bibinfo {author} {\bibfnamefont {Y.}~\bibnamefont {Li}}, \bibinfo {author}
  {\bibfnamefont {A.~D.}\ \bibnamefont {Dunkelberger}}, \bibinfo {author}
  {\bibfnamefont {B.~B.}\ \bibnamefont {Simpkins}}, \bibinfo {author}
  {\bibfnamefont {J.}~\bibnamefont {Yuen-Zhou}},\ and\ \bibinfo {author}
  {\bibfnamefont {W.}~\bibnamefont {Xiong}},\ }\bibfield  {title} {\bibinfo
  {title} {Manipulating optical nonlinearities of molecular polaritons by
  delocalization},\ }\href {https://doi.org/10.1126/sciadv.aax5196} {\bibfield
  {journal} {\bibinfo  {journal} {Science Advances}\ }\textbf {\bibinfo
  {volume} {5}},\ \bibinfo {pages} {aax5196} (\bibinfo {year} {2019})},\
  \Eprint
  {https://arxiv.org/abs/https://advances.sciencemag.org/content/5/9/eaax5196.full.pdf}
  {https://advances.sciencemag.org/content/5/9/eaax5196.full.pdf} \BibitemShut
  {NoStop}%
\bibitem [{\citenamefont {Dunkelberger}\ \emph {et~al.}(2016)\citenamefont
  {Dunkelberger}, \citenamefont {Spann}, \citenamefont {Fears}, \citenamefont
  {Simpkins},\ and\ \citenamefont {Owrutsky}}]{Dunkelberger2016}%
  \BibitemOpen
  \bibfield  {author} {\bibinfo {author} {\bibfnamefont {A.~D.}\ \bibnamefont
  {Dunkelberger}}, \bibinfo {author} {\bibfnamefont {B.~T.}\ \bibnamefont
  {Spann}}, \bibinfo {author} {\bibfnamefont {K.~P.}\ \bibnamefont {Fears}},
  \bibinfo {author} {\bibfnamefont {B.~S.}\ \bibnamefont {Simpkins}},\ and\
  \bibinfo {author} {\bibfnamefont {J.~C.}\ \bibnamefont {Owrutsky}},\
  }\bibfield  {title} {\bibinfo {title} {{Modified relaxation dynamics and
  coherent energy exchange in coupled vibration-cavity polaritons}},\ }\href
  {https://doi.org/10.1038/ncomms13504} {\bibfield  {journal} {\bibinfo
  {journal} {Nature Communications}\ }\textbf {\bibinfo {volume} {7}},\
  \bibinfo {pages} {1} (\bibinfo {year} {2016})}\BibitemShut {NoStop}%
\bibitem [{\citenamefont {Xiang}\ \emph {et~al.}(2018)\citenamefont {Xiang},
  \citenamefont {Ribeiro}, \citenamefont {Dunkelberger}, \citenamefont {Wang},
  \citenamefont {Li}, \citenamefont {Simpkins}, \citenamefont {Owrutsky},
  \citenamefont {Yuen-Zhou},\ and\ \citenamefont {Xiong}}]{Xiang2018}%
  \BibitemOpen
  \bibfield  {author} {\bibinfo {author} {\bibfnamefont {B.}~\bibnamefont
  {Xiang}}, \bibinfo {author} {\bibfnamefont {R.~F.}\ \bibnamefont {Ribeiro}},
  \bibinfo {author} {\bibfnamefont {A.~D.}\ \bibnamefont {Dunkelberger}},
  \bibinfo {author} {\bibfnamefont {J.}~\bibnamefont {Wang}}, \bibinfo {author}
  {\bibfnamefont {Y.}~\bibnamefont {Li}}, \bibinfo {author} {\bibfnamefont
  {B.~S.}\ \bibnamefont {Simpkins}}, \bibinfo {author} {\bibfnamefont {J.~C.}\
  \bibnamefont {Owrutsky}}, \bibinfo {author} {\bibfnamefont {J.}~\bibnamefont
  {Yuen-Zhou}},\ and\ \bibinfo {author} {\bibfnamefont {W.}~\bibnamefont
  {Xiong}},\ }\bibfield  {title} {\bibinfo {title} {Two-dimensional infrared
  spectroscopy of vibrational polaritons},\ }\href
  {https://doi.org/10.1073/pnas.1722063115} {\bibfield  {journal} {\bibinfo
  {journal} {Proceedings of the National Academy of Sciences}\ }\textbf
  {\bibinfo {volume} {115}},\ \bibinfo {pages} {4845} (\bibinfo {year}
  {2018})},\ \Eprint
  {https://arxiv.org/abs/https://www.pnas.org/content/115/19/4845.full.pdf}
  {https://www.pnas.org/content/115/19/4845.full.pdf} \BibitemShut {NoStop}%
\bibitem [{\citenamefont {Dunkelberger}\ \emph
  {et~al.}(2018{\natexlab{a}})\citenamefont {Dunkelberger}, \citenamefont
  {Davidson}, \citenamefont {Ahn}, \citenamefont {Simpkins},\ and\
  \citenamefont {Owrutsky}}]{Dunkelberger2018}%
  \BibitemOpen
  \bibfield  {author} {\bibinfo {author} {\bibfnamefont {A.}~\bibnamefont
  {Dunkelberger}}, \bibinfo {author} {\bibfnamefont {R.}~\bibnamefont
  {Davidson}}, \bibinfo {author} {\bibfnamefont {W.}~\bibnamefont {Ahn}},
  \bibinfo {author} {\bibfnamefont {B.}~\bibnamefont {Simpkins}},\ and\
  \bibinfo {author} {\bibfnamefont {J.}~\bibnamefont {Owrutsky}},\ }\bibfield
  {title} {\bibinfo {title} {{Ultrafast Transmission Modulation and Recovery
  via Vibrational Strong Coupling}},\ }\href
  {https://doi.org/10.1021/acs.jpca.7b10299} {\bibfield  {journal} {\bibinfo
  {journal} {The Journal of Physical Chemistry A}\ }\textbf {\bibinfo {volume}
  {122}},\ \bibinfo {pages} {965} (\bibinfo {year}
  {2018}{\natexlab{a}})}\BibitemShut {NoStop}%
\bibitem [{\citenamefont {Muller}\ \emph {et~al.}(2018)\citenamefont {Muller},
  \citenamefont {Pollard}, \citenamefont {Bechtel}, \citenamefont {Adato},
  \citenamefont {Etezadi}, \citenamefont {Altug},\ and\ \citenamefont
  {Raschke}}]{Muller2018}%
  \BibitemOpen
  \bibfield  {author} {\bibinfo {author} {\bibfnamefont {E.~A.}\ \bibnamefont
  {Muller}}, \bibinfo {author} {\bibfnamefont {B.}~\bibnamefont {Pollard}},
  \bibinfo {author} {\bibfnamefont {H.~A.}\ \bibnamefont {Bechtel}}, \bibinfo
  {author} {\bibfnamefont {R.}~\bibnamefont {Adato}}, \bibinfo {author}
  {\bibfnamefont {D.}~\bibnamefont {Etezadi}}, \bibinfo {author} {\bibfnamefont
  {H.}~\bibnamefont {Altug}},\ and\ \bibinfo {author} {\bibfnamefont {M.~B.}\
  \bibnamefont {Raschke}},\ }\bibfield  {title} {\bibinfo {title} {Nanoimaging
  and control of molecular vibrations through electromagnetically induced
  scattering reaching the strong coupling regime},\ }\bibfield  {booktitle}
  {\emph {\bibinfo {booktitle} {ACS Photonics}},\ }\href
  {https://doi.org/10.1021/acsphotonics.8b00425} {\bibfield  {journal}
  {\bibinfo  {journal} {ACS Photonics}\ }\textbf {\bibinfo {volume} {5}},\
  \bibinfo {pages} {3594} (\bibinfo {year} {2018})}\BibitemShut {NoStop}%
\bibitem [{\citenamefont {Metzger}\ \emph {et~al.}(2019)\citenamefont
  {Metzger}, \citenamefont {Muller}, \citenamefont {Nishida}, \citenamefont
  {Pollard}, \citenamefont {Hentschel},\ and\ \citenamefont
  {Raschke}}]{Metzger2019}%
  \BibitemOpen
  \bibfield  {author} {\bibinfo {author} {\bibfnamefont {B.}~\bibnamefont
  {Metzger}}, \bibinfo {author} {\bibfnamefont {E.}~\bibnamefont {Muller}},
  \bibinfo {author} {\bibfnamefont {J.}~\bibnamefont {Nishida}}, \bibinfo
  {author} {\bibfnamefont {B.}~\bibnamefont {Pollard}}, \bibinfo {author}
  {\bibfnamefont {M.}~\bibnamefont {Hentschel}},\ and\ \bibinfo {author}
  {\bibfnamefont {M.~B.}\ \bibnamefont {Raschke}},\ }\bibfield  {title}
  {\bibinfo {title} {Purcell-enhanced spontaneous emission of molecular
  vibrations},\ }\href {https://doi.org/10.1103/PhysRevLett.123.153001}
  {\bibfield  {journal} {\bibinfo  {journal} {Phys. Rev. Lett.}\ }\textbf
  {\bibinfo {volume} {123}},\ \bibinfo {pages} {153001} (\bibinfo {year}
  {2019})}\BibitemShut {NoStop}%
\bibitem [{\citenamefont {Autore}\ \emph {et~al.}(2018)\citenamefont {Autore},
  \citenamefont {Li}, \citenamefont {Dolado}, \citenamefont {Alfaro-Mozaz},
  \citenamefont {Esteban}, \citenamefont {Atxabal}, \citenamefont {Casanova},
  \citenamefont {Hueso}, \citenamefont {Alonso-Gonz{\'a}lez}, \citenamefont
  {Aizpurua}, \citenamefont {Nikitin}, \citenamefont {V{\'e}lez},\ and\
  \citenamefont {Hillenbrand}}]{Autore:2018}%
  \BibitemOpen
  \bibfield  {author} {\bibinfo {author} {\bibfnamefont {M.}~\bibnamefont
  {Autore}}, \bibinfo {author} {\bibfnamefont {P.}~\bibnamefont {Li}}, \bibinfo
  {author} {\bibfnamefont {I.}~\bibnamefont {Dolado}}, \bibinfo {author}
  {\bibfnamefont {F.~J.}\ \bibnamefont {Alfaro-Mozaz}}, \bibinfo {author}
  {\bibfnamefont {R.}~\bibnamefont {Esteban}}, \bibinfo {author} {\bibfnamefont
  {A.}~\bibnamefont {Atxabal}}, \bibinfo {author} {\bibfnamefont
  {F.}~\bibnamefont {Casanova}}, \bibinfo {author} {\bibfnamefont {L.~E.}\
  \bibnamefont {Hueso}}, \bibinfo {author} {\bibfnamefont {P.}~\bibnamefont
  {Alonso-Gonz{\'a}lez}}, \bibinfo {author} {\bibfnamefont {J.}~\bibnamefont
  {Aizpurua}}, \bibinfo {author} {\bibfnamefont {A.~Y.}\ \bibnamefont
  {Nikitin}}, \bibinfo {author} {\bibfnamefont {S.}~\bibnamefont {V{\'e}lez}},\
  and\ \bibinfo {author} {\bibfnamefont {R.}~\bibnamefont {Hillenbrand}},\
  }\bibfield  {title} {\bibinfo {title} {Boron nitride nanoresonators for
  phonon-enhanced molecular vibrational spectroscopy at the strong coupling
  limit},\ }\href {https://doi.org/10.1038/lsa.2017.172} {\bibfield  {journal}
  {\bibinfo  {journal} {Light: Science \& Applications}\ }\textbf {\bibinfo
  {volume} {7}},\ \bibinfo {pages} {17172} (\bibinfo {year}
  {2018})}\BibitemShut {NoStop}%
\bibitem [{\citenamefont {Folland}\ \emph {et~al.}(2020)\citenamefont
  {Folland}, \citenamefont {Lu}, \citenamefont {Bruncz}, \citenamefont {Nolen},
  \citenamefont {Tadjer},\ and\ \citenamefont {Caldwell}}]{Folland:2020}%
  \BibitemOpen
  \bibfield  {author} {\bibinfo {author} {\bibfnamefont {T.~G.}\ \bibnamefont
  {Folland}}, \bibinfo {author} {\bibfnamefont {G.}~\bibnamefont {Lu}},
  \bibinfo {author} {\bibfnamefont {A.}~\bibnamefont {Bruncz}}, \bibinfo
  {author} {\bibfnamefont {J.~R.}\ \bibnamefont {Nolen}}, \bibinfo {author}
  {\bibfnamefont {M.}~\bibnamefont {Tadjer}},\ and\ \bibinfo {author}
  {\bibfnamefont {J.~D.}\ \bibnamefont {Caldwell}},\ }\bibfield  {title}
  {\bibinfo {title} {Vibrational coupling to epsilon-near-zero waveguide
  modes},\ }\href {https://doi.org/10.1021/acsphotonics.0c00071} {\bibfield
  {journal} {\bibinfo  {journal} {ACS Photonics}\ }\textbf {\bibinfo {volume}
  {7}},\ \bibinfo {pages} {614} (\bibinfo {year} {2020})}\BibitemShut {NoStop}%
\bibitem [{\citenamefont {Kockum}\ \emph {et~al.}(2019)\citenamefont {Kockum},
  \citenamefont {Miranowicz}, \citenamefont {{De Liberato}}, \citenamefont
  {Savasta},\ and\ \citenamefont {Nori}}]{Kockum2019}%
  \BibitemOpen
  \bibfield  {author} {\bibinfo {author} {\bibfnamefont {A.~F.}\ \bibnamefont
  {Kockum}}, \bibinfo {author} {\bibfnamefont {A.}~\bibnamefont {Miranowicz}},
  \bibinfo {author} {\bibfnamefont {S.}~\bibnamefont {{De Liberato}}}, \bibinfo
  {author} {\bibfnamefont {S.}~\bibnamefont {Savasta}},\ and\ \bibinfo {author}
  {\bibfnamefont {F.}~\bibnamefont {Nori}},\ }\bibfield  {title} {\bibinfo
  {title} {{Ultrastrong coupling between light and matter}},\ }\href
  {https://doi.org/10.1038/s42254-018-0006-2} {\bibfield  {journal} {\bibinfo
  {journal} {Nature Reviews Physics}\ }\textbf {\bibinfo {volume} {1}},\
  \bibinfo {pages} {19} (\bibinfo {year} {2019})},\ \Eprint
  {https://arxiv.org/abs/1807.11636} {1807.11636} \BibitemShut {NoStop}%
\bibitem [{\citenamefont {Forn-D{\'{i}}az}\ \emph {et~al.}(2019)\citenamefont
  {Forn-D{\'{i}}az}, \citenamefont {Lamata}, \citenamefont {Rico},
  \citenamefont {Kono},\ and\ \citenamefont {Solano}}]{Forn-Diaz2018}%
  \BibitemOpen
  \bibfield  {author} {\bibinfo {author} {\bibfnamefont {P.}~\bibnamefont
  {Forn-D{\'{i}}az}}, \bibinfo {author} {\bibfnamefont {L.}~\bibnamefont
  {Lamata}}, \bibinfo {author} {\bibfnamefont {E.}~\bibnamefont {Rico}},
  \bibinfo {author} {\bibfnamefont {J.}~\bibnamefont {Kono}},\ and\ \bibinfo
  {author} {\bibfnamefont {E.}~\bibnamefont {Solano}},\ }\bibfield  {title}
  {\bibinfo {title} {{Ultrastrong coupling regimes of light-matter
  interaction}},\ }\href@noop {} {\bibfield  {journal} {\bibinfo  {journal}
  {Rev. Mod. Phys.}\ }\textbf {\bibinfo {volume} {91}},\ \bibinfo {pages}
  {25005} (\bibinfo {year} {2019})}\BibitemShut {NoStop}%
\bibitem [{\citenamefont {George}\ \emph {et~al.}(2016)\citenamefont {George},
  \citenamefont {Chervy}, \citenamefont {Shalabney}, \citenamefont {Devaux},
  \citenamefont {Hiura}, \citenamefont {Genet},\ and\ \citenamefont
  {Ebbesen}}]{George2016}%
  \BibitemOpen
  \bibfield  {author} {\bibinfo {author} {\bibfnamefont {J.}~\bibnamefont
  {George}}, \bibinfo {author} {\bibfnamefont {T.}~\bibnamefont {Chervy}},
  \bibinfo {author} {\bibfnamefont {A.}~\bibnamefont {Shalabney}}, \bibinfo
  {author} {\bibfnamefont {E.}~\bibnamefont {Devaux}}, \bibinfo {author}
  {\bibfnamefont {H.}~\bibnamefont {Hiura}}, \bibinfo {author} {\bibfnamefont
  {C.}~\bibnamefont {Genet}},\ and\ \bibinfo {author} {\bibfnamefont {T.~W.}\
  \bibnamefont {Ebbesen}},\ }\bibfield  {title} {\bibinfo {title} {{Multiple
  Rabi Splittings under Ultrastrong Vibrational Coupling}},\ }\href
  {https://doi.org/10.1103/PhysRevLett.117.153601} {\bibfield  {journal}
  {\bibinfo  {journal} {Physical Review Letters}\ }\textbf {\bibinfo {volume}
  {117}},\ \bibinfo {pages} {153601} (\bibinfo {year} {2016})}\BibitemShut
  {NoStop}%
\bibitem [{\citenamefont {Askenazi}\ \emph {et~al.}(2017)\citenamefont
  {Askenazi}, \citenamefont {Vasanelli}, \citenamefont {Todorov}, \citenamefont
  {Sakat}, \citenamefont {Greffet}, \citenamefont {Beaudoin}, \citenamefont
  {Sagnes},\ and\ \citenamefont {Sirtori}}]{Askenazi2017}%
  \BibitemOpen
  \bibfield  {author} {\bibinfo {author} {\bibfnamefont {B.}~\bibnamefont
  {Askenazi}}, \bibinfo {author} {\bibfnamefont {A.}~\bibnamefont {Vasanelli}},
  \bibinfo {author} {\bibfnamefont {Y.}~\bibnamefont {Todorov}}, \bibinfo
  {author} {\bibfnamefont {E.}~\bibnamefont {Sakat}}, \bibinfo {author}
  {\bibfnamefont {J.-J.}\ \bibnamefont {Greffet}}, \bibinfo {author}
  {\bibfnamefont {G.}~\bibnamefont {Beaudoin}}, \bibinfo {author}
  {\bibfnamefont {I.}~\bibnamefont {Sagnes}},\ and\ \bibinfo {author}
  {\bibfnamefont {C.}~\bibnamefont {Sirtori}},\ }\bibfield  {title} {\bibinfo
  {title} {Midinfrared ultrastrong light--matter coupling for thz thermal
  emission},\ }\bibfield  {booktitle} {\emph {\bibinfo {booktitle} {ACS
  Photonics}},\ }\href {https://doi.org/10.1021/acsphotonics.7b00838}
  {\bibfield  {journal} {\bibinfo  {journal} {ACS Photonics}\ }\textbf
  {\bibinfo {volume} {4}},\ \bibinfo {pages} {2550} (\bibinfo {year}
  {2017})}\BibitemShut {NoStop}%
\bibitem [{\citenamefont {Yoo}\ \emph {et~al.}(2021)\citenamefont {Yoo},
  \citenamefont {de~Le{\'o}n-P{\'e}rez}, \citenamefont {Pelton}, \citenamefont
  {Lee}, \citenamefont {Mohr}, \citenamefont {Raschke}, \citenamefont
  {Caldwell}, \citenamefont {Mart{\'\i}n-Moreno},\ and\ \citenamefont
  {Oh}}]{Yoo:2021}%
  \BibitemOpen
  \bibfield  {author} {\bibinfo {author} {\bibfnamefont {D.}~\bibnamefont
  {Yoo}}, \bibinfo {author} {\bibfnamefont {F.}~\bibnamefont
  {de~Le{\'o}n-P{\'e}rez}}, \bibinfo {author} {\bibfnamefont {M.}~\bibnamefont
  {Pelton}}, \bibinfo {author} {\bibfnamefont {I.-H.}\ \bibnamefont {Lee}},
  \bibinfo {author} {\bibfnamefont {D.~A.}\ \bibnamefont {Mohr}}, \bibinfo
  {author} {\bibfnamefont {M.~B.}\ \bibnamefont {Raschke}}, \bibinfo {author}
  {\bibfnamefont {J.~D.}\ \bibnamefont {Caldwell}}, \bibinfo {author}
  {\bibfnamefont {L.}~\bibnamefont {Mart{\'\i}n-Moreno}},\ and\ \bibinfo
  {author} {\bibfnamefont {S.-H.}\ \bibnamefont {Oh}},\ }\bibfield  {title}
  {\bibinfo {title} {Ultrastrong plasmon--phonon coupling via epsilon-near-zero
  nanocavities},\ }\href {https://doi.org/10.1038/s41566-020-00731-5}
  {\bibfield  {journal} {\bibinfo  {journal} {Nature Photonics}\ }\textbf
  {\bibinfo {volume} {15}},\ \bibinfo {pages} {125} (\bibinfo {year}
  {2021})}\BibitemShut {NoStop}%
\bibitem [{\citenamefont {Dodonov}(2020)}]{Dodonov2020fifty}%
  \BibitemOpen
  \bibfield  {author} {\bibinfo {author} {\bibfnamefont {V.}~\bibnamefont
  {Dodonov}},\ }\bibfield  {title} {\bibinfo {title} {Fifty years of the
  dynamical casimir effect},\ }\href@noop {} {\bibfield  {journal} {\bibinfo
  {journal} {Physics}\ }\textbf {\bibinfo {volume} {2}},\ \bibinfo {pages} {67}
  (\bibinfo {year} {2020})}\BibitemShut {NoStop}%
\bibitem [{\citenamefont {Jun}\ \emph {et~al.}(2012)\citenamefont {Jun},
  \citenamefont {Gonzales}, \citenamefont {Reno}, \citenamefont {Shaner},
  \citenamefont {Gabbay},\ and\ \citenamefont {Brener}}]{Jun2012}%
  \BibitemOpen
  \bibfield  {author} {\bibinfo {author} {\bibfnamefont {Y.~C.}\ \bibnamefont
  {Jun}}, \bibinfo {author} {\bibfnamefont {E.}~\bibnamefont {Gonzales}},
  \bibinfo {author} {\bibfnamefont {J.~L.}\ \bibnamefont {Reno}}, \bibinfo
  {author} {\bibfnamefont {E.~A.}\ \bibnamefont {Shaner}}, \bibinfo {author}
  {\bibfnamefont {A.}~\bibnamefont {Gabbay}},\ and\ \bibinfo {author}
  {\bibfnamefont {I.}~\bibnamefont {Brener}},\ }\bibfield  {title} {\bibinfo
  {title} {Active tuning of mid-infrared metamaterials by electrical control of
  carrier densities},\ }\href {https://doi.org/10.1364/OE.20.001903} {\bibfield
   {journal} {\bibinfo  {journal} {Opt. Express}\ }\textbf {\bibinfo {volume}
  {20}},\ \bibinfo {pages} {1903} (\bibinfo {year} {2012})}\BibitemShut
  {NoStop}%
\bibitem [{\citenamefont {Panah}\ \emph {et~al.}(2017)\citenamefont {Panah},
  \citenamefont {Han}, \citenamefont {Norrman}, \citenamefont {Pryds},
  \citenamefont {Nadtochiy}, \citenamefont {Zhukov}, \citenamefont
  {Lavrinenko},\ and\ \citenamefont {Semenova}}]{Panah2017}%
  \BibitemOpen
  \bibfield  {author} {\bibinfo {author} {\bibfnamefont {M.~E.~A.}\
  \bibnamefont {Panah}}, \bibinfo {author} {\bibfnamefont {L.}~\bibnamefont
  {Han}}, \bibinfo {author} {\bibfnamefont {K.}~\bibnamefont {Norrman}},
  \bibinfo {author} {\bibfnamefont {N.}~\bibnamefont {Pryds}}, \bibinfo
  {author} {\bibfnamefont {A.}~\bibnamefont {Nadtochiy}}, \bibinfo {author}
  {\bibfnamefont {A.}~\bibnamefont {Zhukov}}, \bibinfo {author} {\bibfnamefont
  {A.~V.}\ \bibnamefont {Lavrinenko}},\ and\ \bibinfo {author} {\bibfnamefont
  {E.~S.}\ \bibnamefont {Semenova}},\ }\bibfield  {title} {\bibinfo {title}
  {Mid-ir optical properties of silicon doped inp},\ }\href
  {https://doi.org/10.1364/OME.7.002260} {\bibfield  {journal} {\bibinfo
  {journal} {Opt. Mater. Express}\ }\textbf {\bibinfo {volume} {7}},\ \bibinfo
  {pages} {2260} (\bibinfo {year} {2017})}\BibitemShut {NoStop}%
\bibitem [{\citenamefont {Dunkelberger}\ \emph
  {et~al.}(2018{\natexlab{b}})\citenamefont {Dunkelberger}, \citenamefont
  {Ellis}, \citenamefont {Ratchford}, \citenamefont {Giles}, \citenamefont
  {Kim}, \citenamefont {Kim}, \citenamefont {Spann}, \citenamefont
  {Vurgaftman}, \citenamefont {Tischler}, \citenamefont {Long}, \citenamefont
  {Glembocki}, \citenamefont {Owrutsky},\ and\ \citenamefont
  {Caldwell}}]{Dunkelberger:2018}%
  \BibitemOpen
  \bibfield  {author} {\bibinfo {author} {\bibfnamefont {A.~D.}\ \bibnamefont
  {Dunkelberger}}, \bibinfo {author} {\bibfnamefont {C.~T.}\ \bibnamefont
  {Ellis}}, \bibinfo {author} {\bibfnamefont {D.~C.}\ \bibnamefont
  {Ratchford}}, \bibinfo {author} {\bibfnamefont {A.~J.}\ \bibnamefont
  {Giles}}, \bibinfo {author} {\bibfnamefont {M.}~\bibnamefont {Kim}}, \bibinfo
  {author} {\bibfnamefont {C.~S.}\ \bibnamefont {Kim}}, \bibinfo {author}
  {\bibfnamefont {B.~T.}\ \bibnamefont {Spann}}, \bibinfo {author}
  {\bibfnamefont {I.}~\bibnamefont {Vurgaftman}}, \bibinfo {author}
  {\bibfnamefont {J.~G.}\ \bibnamefont {Tischler}}, \bibinfo {author}
  {\bibfnamefont {J.~P.}\ \bibnamefont {Long}}, \bibinfo {author}
  {\bibfnamefont {O.~J.}\ \bibnamefont {Glembocki}}, \bibinfo {author}
  {\bibfnamefont {J.~C.}\ \bibnamefont {Owrutsky}},\ and\ \bibinfo {author}
  {\bibfnamefont {J.~D.}\ \bibnamefont {Caldwell}},\ }\bibfield  {title}
  {\bibinfo {title} {Active tuning of surface phonon polariton resonances via
  carrier photoinjection},\ }\href {https://doi.org/10.1038/s41566-017-0069-0}
  {\bibfield  {journal} {\bibinfo  {journal} {Nature Photonics}\ }\textbf
  {\bibinfo {volume} {12}},\ \bibinfo {pages} {50} (\bibinfo {year}
  {2018}{\natexlab{b}})}\BibitemShut {NoStop}%
\bibitem [{\citenamefont {Dunkelberger}\ \emph {et~al.}(2020)\citenamefont
  {Dunkelberger}, \citenamefont {Ratchford}, \citenamefont {Grafton},
  \citenamefont {Breslin}, \citenamefont {Ryland}, \citenamefont {Katzer},
  \citenamefont {Fears}, \citenamefont {Weiblen}, \citenamefont {Vurgaftman},
  \citenamefont {Giles}, \citenamefont {Ellis}, \citenamefont {Tischler},
  \citenamefont {Caldwell},\ and\ \citenamefont
  {Owrutsky}}]{Dunkelberger2020tuning}%
  \BibitemOpen
  \bibfield  {author} {\bibinfo {author} {\bibfnamefont {A.~D.}\ \bibnamefont
  {Dunkelberger}}, \bibinfo {author} {\bibfnamefont {D.~C.}\ \bibnamefont
  {Ratchford}}, \bibinfo {author} {\bibfnamefont {A.~B.}\ \bibnamefont
  {Grafton}}, \bibinfo {author} {\bibfnamefont {V.~M.}\ \bibnamefont
  {Breslin}}, \bibinfo {author} {\bibfnamefont {E.~S.}\ \bibnamefont {Ryland}},
  \bibinfo {author} {\bibfnamefont {D.~S.}\ \bibnamefont {Katzer}}, \bibinfo
  {author} {\bibfnamefont {K.~P.}\ \bibnamefont {Fears}}, \bibinfo {author}
  {\bibfnamefont {R.~J.}\ \bibnamefont {Weiblen}}, \bibinfo {author}
  {\bibfnamefont {I.}~\bibnamefont {Vurgaftman}}, \bibinfo {author}
  {\bibfnamefont {A.~J.}\ \bibnamefont {Giles}}, \bibinfo {author}
  {\bibfnamefont {C.~T.}\ \bibnamefont {Ellis}}, \bibinfo {author}
  {\bibfnamefont {J.~G.}\ \bibnamefont {Tischler}}, \bibinfo {author}
  {\bibfnamefont {J.~D.}\ \bibnamefont {Caldwell}},\ and\ \bibinfo {author}
  {\bibfnamefont {J.~C.}\ \bibnamefont {Owrutsky}},\ }\bibfield  {title}
  {\bibinfo {title} {Ultrafast active tuning of the berreman mode},\ }\bibfield
   {booktitle} {\emph {\bibinfo {booktitle} {ACS Photonics}},\ }\href
  {https://doi.org/10.1021/acsphotonics.9b01578} {\bibfield  {journal}
  {\bibinfo  {journal} {ACS Photonics}\ }\textbf {\bibinfo {volume} {7}},\
  \bibinfo {pages} {279} (\bibinfo {year} {2020})}\BibitemShut {NoStop}%
\bibitem [{\citenamefont {Hern{\'{a}}ndez}\ and\ \citenamefont
  {Herrera}(2019)}]{Hernandez2019}%
  \BibitemOpen
  \bibfield  {author} {\bibinfo {author} {\bibfnamefont {F.}~\bibnamefont
  {Hern{\'{a}}ndez}}\ and\ \bibinfo {author} {\bibfnamefont {F.}~\bibnamefont
  {Herrera}},\ }\bibfield  {title} {\bibinfo {title} {{Multi-level quantum Rabi
  model for anharmonic vibrational polaritons}},\ }\href
  {https://doi.org/10.1063/1.5121426} {\bibfield  {journal} {\bibinfo
  {journal} {The Journal of Chemical Physics}\ }\textbf {\bibinfo {volume}
  {151}},\ \bibinfo {pages} {144116} (\bibinfo {year} {2019})}\BibitemShut
  {NoStop}%
\bibitem [{\citenamefont {Triana}\ \emph {et~al.}(2020)\citenamefont {Triana},
  \citenamefont {Hern{\'{a}}ndez},\ and\ \citenamefont {Herrera}}]{Triana2020}%
  \BibitemOpen
  \bibfield  {author} {\bibinfo {author} {\bibfnamefont {J.}~\bibnamefont
  {Triana}}, \bibinfo {author} {\bibfnamefont {F.}~\bibnamefont
  {Hern{\'{a}}ndez}},\ and\ \bibinfo {author} {\bibfnamefont {F.}~\bibnamefont
  {Herrera}},\ }\bibfield  {title} {\bibinfo {title} {{The shape of the
  electric dipole function determines the sub-picosecond dynamics of anharmonic
  vibrational polaritons}},\ }\href {https://doi.org/10.1063/5.0009869}
  {\bibfield  {journal} {\bibinfo  {journal} {Journal of Chemical Physics}\
  }\textbf {\bibinfo {volume} {152}},\ \bibinfo {pages} {234111} (\bibinfo
  {year} {2020})}\BibitemShut {NoStop}%
\bibitem [{\citenamefont {Andrews}\ \emph {et~al.}(2018)\citenamefont
  {Andrews}, \citenamefont {Jones}, \citenamefont {Salam},\ and\ \citenamefont
  {Woolley}}]{Andrews2018}%
  \BibitemOpen
  \bibfield  {author} {\bibinfo {author} {\bibfnamefont {D.~L.}\ \bibnamefont
  {Andrews}}, \bibinfo {author} {\bibfnamefont {G.~A.}\ \bibnamefont {Jones}},
  \bibinfo {author} {\bibfnamefont {A.}~\bibnamefont {Salam}},\ and\ \bibinfo
  {author} {\bibfnamefont {R.~G.}\ \bibnamefont {Woolley}},\ }\bibfield
  {title} {\bibinfo {title} {{Perspective: Quantum Hamiltonians for optical
  interactions}},\ }\href {https://doi.org/10.1063/1.5018399} {\bibfield
  {journal} {\bibinfo  {journal} {The Journal of Chemical Physics}\ }\textbf
  {\bibinfo {volume} {148}},\ \bibinfo {pages} {40901} (\bibinfo {year}
  {2018})}\BibitemShut {NoStop}%
\bibitem [{\citenamefont {Elsaesser}(1991)}]{Elsaesser1991}%
  \BibitemOpen
  \bibfield  {author} {\bibinfo {author} {\bibfnamefont {T.}~\bibnamefont
  {Elsaesser}},\ }\bibfield  {title} {\bibinfo {title} {{Vibrational And
  Vibronic Relaxation Of Large Polyatomic Molecules In Liquids}},\ }\href
  {https://doi.org/10.1146/annurev.physchem.42.1.83} {\bibfield  {journal}
  {\bibinfo  {journal} {Annual Review of Physical Chemistry}\ }\textbf
  {\bibinfo {volume} {42}},\ \bibinfo {pages} {83} (\bibinfo {year}
  {1991})}\BibitemShut {NoStop}%
\bibitem [{\citenamefont {Morse}(1929)}]{Morse1929}%
  \BibitemOpen
  \bibfield  {author} {\bibinfo {author} {\bibfnamefont {P.~M.}\ \bibnamefont
  {Morse}},\ }\bibfield  {title} {\bibinfo {title} {{Diatomic Molecules
  According to the Wave Mechanics. II. Vibrational Levels}},\ }\href
  {https://doi.org/10.1103/PhysRev.34.57} {\bibfield  {journal} {\bibinfo
  {journal} {Physical Review}\ }\textbf {\bibinfo {volume} {34}},\ \bibinfo
  {pages} {57} (\bibinfo {year} {1929})}\BibitemShut {NoStop}%
\bibitem [{\citenamefont {{F. Ribeiro}}\ \emph {et~al.}(2018)\citenamefont {{F.
  Ribeiro}}, \citenamefont {Dunkelberger}, \citenamefont {Xiang}, \citenamefont
  {Xiong}, \citenamefont {Simpkins}, \citenamefont {Owrutsky},\ and\
  \citenamefont {Yuen-Zhou}}]{Ribeiro2018}%
  \BibitemOpen
  \bibfield  {author} {\bibinfo {author} {\bibfnamefont {R.}~\bibnamefont {{F.
  Ribeiro}}}, \bibinfo {author} {\bibfnamefont {A.~D.}\ \bibnamefont
  {Dunkelberger}}, \bibinfo {author} {\bibfnamefont {B.}~\bibnamefont {Xiang}},
  \bibinfo {author} {\bibfnamefont {W.}~\bibnamefont {Xiong}}, \bibinfo
  {author} {\bibfnamefont {B.~S.}\ \bibnamefont {Simpkins}}, \bibinfo {author}
  {\bibfnamefont {J.~C.}\ \bibnamefont {Owrutsky}},\ and\ \bibinfo {author}
  {\bibfnamefont {J.}~\bibnamefont {Yuen-Zhou}},\ }\bibfield  {title} {\bibinfo
  {title} {{Theory for Nonlinear Spectroscopy of Vibrational Polaritons}},\
  }\href {https://doi.org/10.1021/acs.jpclett.8b01176} {\bibfield  {journal}
  {\bibinfo  {journal} {The Journal of Physical Chemistry Letters}\ }\textbf
  {\bibinfo {volume} {9}},\ \bibinfo {pages} {3766} (\bibinfo {year}
  {2018})}\BibitemShut {NoStop}%
\bibitem [{\citenamefont {Grafton}\ \emph
  {et~al.}(2021{\natexlab{b}})\citenamefont {Grafton}, \citenamefont
  {Dunkelberger}, \citenamefont {Simpkins}, \citenamefont {Triana},
  \citenamefont {Hern{\'{a}}ndez}, \citenamefont {Herrera},\ and\ \citenamefont
  {Owrutsky}}]{Grafton2021}%
  \BibitemOpen
  \bibfield  {author} {\bibinfo {author} {\bibfnamefont {A.~B.}\ \bibnamefont
  {Grafton}}, \bibinfo {author} {\bibfnamefont {A.~D.}\ \bibnamefont
  {Dunkelberger}}, \bibinfo {author} {\bibfnamefont {B.~S.}\ \bibnamefont
  {Simpkins}}, \bibinfo {author} {\bibfnamefont {J.~F.}\ \bibnamefont
  {Triana}}, \bibinfo {author} {\bibfnamefont {F.~J.}\ \bibnamefont
  {Hern{\'{a}}ndez}}, \bibinfo {author} {\bibfnamefont {F.}~\bibnamefont
  {Herrera}},\ and\ \bibinfo {author} {\bibfnamefont {J.~C.}\ \bibnamefont
  {Owrutsky}},\ }\bibfield  {title} {\bibinfo {title} {{Excited-state
  vibration-polariton transitions and dynamics in nitroprusside}},\ }\href
  {https://doi.org/10.1038/s41467-020-20535-z} {\bibfield  {journal} {\bibinfo
  {journal} {Nature Communications}\ }\textbf {\bibinfo {volume} {12}},\
  \bibinfo {pages} {214} (\bibinfo {year} {2021}{\natexlab{b}})}\BibitemShut
  {NoStop}%
\bibitem [{\citenamefont {Triana}\ \emph {et~al.}(2018)\citenamefont {Triana},
  \citenamefont {Pel{\'{a}}ez},\ and\ \citenamefont
  {Sanz-Vicario}}]{Triana2018}%
  \BibitemOpen
  \bibfield  {author} {\bibinfo {author} {\bibfnamefont {J.}~\bibnamefont
  {Triana}}, \bibinfo {author} {\bibfnamefont {D.}~\bibnamefont
  {Pel{\'{a}}ez}},\ and\ \bibinfo {author} {\bibfnamefont {J.}~\bibnamefont
  {Sanz-Vicario}},\ }\bibfield  {title} {\bibinfo {title} {{Entangled
  Photonic-Nuclear Molecular Dynamics of LiF in Quantum Optical Cavities}},\
  }\bibfield  {journal} {\bibinfo  {journal} {Journal of Physical Chemistry A}\
  }\textbf {\bibinfo {volume} {122}},\ \href
  {https://doi.org/10.1021/acs.jpca.7b11833} {10.1021/acs.jpca.7b11833}
  (\bibinfo {year} {2018})\BibitemShut {NoStop}%
\bibitem [{\citenamefont {Beck}\ \emph {et~al.}(2000)\citenamefont {Beck},
  \citenamefont {Jackle}, \citenamefont {Worth},\ and\ \citenamefont
  {Meyer}}]{mctdhpaper}%
  \BibitemOpen
  \bibfield  {author} {\bibinfo {author} {\bibfnamefont {M.~H.}\ \bibnamefont
  {Beck}}, \bibinfo {author} {\bibfnamefont {A.}~\bibnamefont {Jackle}},
  \bibinfo {author} {\bibfnamefont {G.~A.}\ \bibnamefont {Worth}},\ and\
  \bibinfo {author} {\bibfnamefont {H.-D.}\ \bibnamefont {Meyer}},\ }\bibfield
  {title} {\bibinfo {title} {{The multiconfiguration time-dependent Hartree
  (MCTDH) method: a highly efficient algorithm for propagating wavepackets}},\
  }\href {https://doi.org/https://doi.org/10.1016/S0370-1573(99)00047-2}
  {\bibfield  {journal} {\bibinfo  {journal} {Physics Reports}\ }\textbf
  {\bibinfo {volume} {324}},\ \bibinfo {pages} {1} (\bibinfo {year}
  {2000})}\BibitemShut {NoStop}%
\bibitem [{\citenamefont {Meyer}\ \emph {et~al.}(2009)\citenamefont {Meyer},
  \citenamefont {Gatti},\ and\ \citenamefont {Worth}}]{mctdhbook}%
  \BibitemOpen
  \bibfield  {author} {\bibinfo {author} {\bibfnamefont {H.~D.}\ \bibnamefont
  {Meyer}}, \bibinfo {author} {\bibfnamefont {F.}~\bibnamefont {Gatti}},\ and\
  \bibinfo {author} {\bibfnamefont {G.~A.}\ \bibnamefont {Worth}},\ }\href
  {https://books.google.fr/books?id=LMVLvheAeEAC} {\emph {\bibinfo {title}
  {{Multidimensional {Q}uantum {D}ynamics: {MCTDH} {T}heory and
  {a}pplications}}}}\ (\bibinfo  {publisher} {John Wiley {\&} Sons},\ \bibinfo
  {year} {2009})\BibitemShut {NoStop}%
\bibitem [{\citenamefont {Vendrell}\ \emph {et~al.}(2007)\citenamefont
  {Vendrell}, \citenamefont {Gatti},\ and\ \citenamefont
  {Meyer}}]{Vendrell2007}%
  \BibitemOpen
  \bibfield  {author} {\bibinfo {author} {\bibfnamefont {O.}~\bibnamefont
  {Vendrell}}, \bibinfo {author} {\bibfnamefont {F.}~\bibnamefont {Gatti}},\
  and\ \bibinfo {author} {\bibfnamefont {H.-D.}\ \bibnamefont {Meyer}},\
  }\bibfield  {title} {\bibinfo {title} {Full dimensional (15-dimensional)
  quantum-dynamical simulation of the protonated water dimer. ii. infrared
  spectrum and vibrational dynamics},\ }\href
  {https://doi.org/10.1063/1.2787596} {\bibfield  {journal} {\bibinfo
  {journal} {The Journal of Chemical Physics}\ }\textbf {\bibinfo {volume}
  {127}},\ \bibinfo {pages} {184303} (\bibinfo {year} {2007})},\ \Eprint
  {https://arxiv.org/abs/https://doi.org/10.1063/1.2787596}
  {https://doi.org/10.1063/1.2787596} \BibitemShut {NoStop}%
\bibitem [{\citenamefont {Vendrell}(2018)}]{Vendrell2018}%
  \BibitemOpen
  \bibfield  {author} {\bibinfo {author} {\bibfnamefont {O.}~\bibnamefont
  {Vendrell}},\ }\bibfield  {title} {\bibinfo {title} {Collective jahn-teller
  interactions through light-matter coupling in a cavity},\ }\href
  {https://doi.org/10.1103/PhysRevLett.121.253001} {\bibfield  {journal}
  {\bibinfo  {journal} {Phys. Rev. Lett.}\ }\textbf {\bibinfo {volume} {121}},\
  \bibinfo {pages} {253001} (\bibinfo {year} {2018})}\BibitemShut {NoStop}%
\bibitem [{\citenamefont {Triana}\ and\ \citenamefont
  {Herrera}(2020)}]{Triana2020sd}%
  \BibitemOpen
  \bibfield  {author} {\bibinfo {author} {\bibfnamefont {J.}~\bibnamefont
  {Triana}}\ and\ \bibinfo {author} {\bibfnamefont {F.}~\bibnamefont
  {Herrera}},\ }\bibfield  {title} {\bibinfo {title} {{Self-Dissociation of
  Polar Molecules in a Confined Infrared Vacuum}},\ }\bibfield  {journal}
  {\bibinfo  {journal} {ChemXiv:12702419}\ }\href
  {https://doi.org/10.26434/chemrxiv.12702419.v1}
  {10.26434/chemrxiv.12702419.v1} (\bibinfo {year} {2020}),\ \Eprint
  {https://arxiv.org/abs/12702419} {ChemXiv:12702419} \BibitemShut {NoStop}%
\bibitem [{\citenamefont {Meyer}\ and\ \citenamefont
  {Worth}(2003)}]{Meyer2003}%
  \BibitemOpen
  \bibfield  {author} {\bibinfo {author} {\bibfnamefont {H.-D.}\ \bibnamefont
  {Meyer}}\ and\ \bibinfo {author} {\bibfnamefont {G.~A.}\ \bibnamefont
  {Worth}},\ }\bibfield  {title} {\bibinfo {title} {{Quantum molecular
  dynamics: propagating wavepackets and density operators using the
  multiconfiguration time-dependent Hartree method}},\ }\href
  {https://doi.org/10.1007/s00214-003-0439-1} {\bibfield  {journal} {\bibinfo
  {journal} {Theoretical Chemistry Accounts}\ }\textbf {\bibinfo {volume}
  {109}},\ \bibinfo {pages} {251} (\bibinfo {year} {2003})}\BibitemShut
  {NoStop}%
\bibitem [{\citenamefont {Meyer}\ \emph {et~al.}(2006)\citenamefont {Meyer},
  \citenamefont {Qu{\'e}r{\'e}}, \citenamefont {L{\'e}onard},\ and\
  \citenamefont {Gatti}}]{Meyer2006}%
  \BibitemOpen
  \bibfield  {author} {\bibinfo {author} {\bibfnamefont {H.-D.}\ \bibnamefont
  {Meyer}}, \bibinfo {author} {\bibfnamefont {F.~L.}\ \bibnamefont
  {Qu{\'e}r{\'e}}}, \bibinfo {author} {\bibfnamefont {C.}~\bibnamefont
  {L{\'e}onard}},\ and\ \bibinfo {author} {\bibfnamefont {F.}~\bibnamefont
  {Gatti}},\ }\bibfield  {title} {\bibinfo {title} {Calculation and selective
  population of vibrational levels with the multiconfiguration time-dependent
  hartree (mctdh) algorithm},\ }\href
  {https://doi.org/https://doi.org/10.1016/j.chemphys.2006.06.002} {\bibfield
  {journal} {\bibinfo  {journal} {Chemical Physics}\ }\textbf {\bibinfo
  {volume} {329}},\ \bibinfo {pages} {179} (\bibinfo {year} {2006})},\ \bibinfo
  {note} {electron Correlation and Multimode Dynamics in Molecules}\BibitemShut
  {NoStop}%
\bibitem [{\citenamefont {Barnett}\ and\ \citenamefont
  {Radmore}(1997)}]{Barnett-Radmore}%
  \BibitemOpen
  \bibfield  {author} {\bibinfo {author} {\bibfnamefont {S.~M.}\ \bibnamefont
  {Barnett}}\ and\ \bibinfo {author} {\bibfnamefont {P.}~\bibnamefont
  {Radmore}},\ }\href@noop {} {\emph {\bibinfo {title} {Methods in theoretical
  quantum optics}}}\ (\bibinfo  {publisher} {Oxford University Press},\
  \bibinfo {year} {1997})\BibitemShut {NoStop}%
\bibitem [{\citenamefont {Gerry}\ and\ \citenamefont
  {Knight}(2005)}]{Gerry2005}%
  \BibitemOpen
  \bibfield  {author} {\bibinfo {author} {\bibfnamefont {C.}~\bibnamefont
  {Gerry}}\ and\ \bibinfo {author} {\bibfnamefont {P.}~\bibnamefont {Knight}},\
  }\href {https://books.google.com.co/books?id=CgByyoBJJwgC} {\emph {\bibinfo
  {title} {{Introductory quantum optics}}}}\ (\bibinfo  {publisher} {Cambridge
  University Press, Cambridge},\ \bibinfo {year} {2005})\BibitemShut {NoStop}%
\bibitem [{\citenamefont {Mandel}(1982)}]{Mandel1982}%
  \BibitemOpen
  \bibfield  {author} {\bibinfo {author} {\bibfnamefont {L.}~\bibnamefont
  {Mandel}},\ }\bibfield  {title} {\bibinfo {title} {Squeezed states and
  sub-poissonian photon statistics},\ }\href
  {https://doi.org/10.1103/PhysRevLett.49.136} {\bibfield  {journal} {\bibinfo
  {journal} {Phys. Rev. Lett.}\ }\textbf {\bibinfo {volume} {49}},\ \bibinfo
  {pages} {136} (\bibinfo {year} {1982})}\BibitemShut {NoStop}%
\bibitem [{\citenamefont {Andersen}\ \emph {et~al.}(2016)\citenamefont
  {Andersen}, \citenamefont {Gehring}, \citenamefont {Marquardt},\ and\
  \citenamefont {Leuchs}}]{Andersen2016}%
  \BibitemOpen
  \bibfield  {author} {\bibinfo {author} {\bibfnamefont {U.~L.}\ \bibnamefont
  {Andersen}}, \bibinfo {author} {\bibfnamefont {T.}~\bibnamefont {Gehring}},
  \bibinfo {author} {\bibfnamefont {C.}~\bibnamefont {Marquardt}},\ and\
  \bibinfo {author} {\bibfnamefont {G.}~\bibnamefont {Leuchs}},\ }\bibfield
  {title} {\bibinfo {title} {30 years of squeezed light generation},\ }\href
  {https://doi.org/10.1088/0031-8949/91/5/053001} {\bibfield  {journal}
  {\bibinfo  {journal} {Physica Scripta}\ }\textbf {\bibinfo {volume} {91}},\
  \bibinfo {pages} {053001} (\bibinfo {year} {2016})}\BibitemShut {NoStop}%
\bibitem [{\citenamefont {Sarandy}\ \emph {et~al.}(2004)\citenamefont
  {Sarandy}, \citenamefont {Wu},\ and\ \citenamefont {Lidar}}]{Sarandy2004}%
  \BibitemOpen
  \bibfield  {author} {\bibinfo {author} {\bibfnamefont {M.~S.}\ \bibnamefont
  {Sarandy}}, \bibinfo {author} {\bibfnamefont {L.~A.}\ \bibnamefont {Wu}},\
  and\ \bibinfo {author} {\bibfnamefont {D.~A.}\ \bibnamefont {Lidar}},\
  }\bibfield  {title} {\bibinfo {title} {Consistency of the adiabatic
  theorem},\ }\href {https://doi.org/10.1007/s11128-004-7712-7} {\bibfield
  {journal} {\bibinfo  {journal} {Quantum Information Processing}\ }\textbf
  {\bibinfo {volume} {3}},\ \bibinfo {pages} {331} (\bibinfo {year}
  {2004})}\BibitemShut {NoStop}%
\bibitem [{\citenamefont {Ji}\ \emph {et~al.}(2018)\citenamefont {Ji},
  \citenamefont {Zhang}, \citenamefont {Wang}, \citenamefont {Lv},\ and\
  \citenamefont {Shen}}]{Ji2018}%
  \BibitemOpen
  \bibfield  {author} {\bibinfo {author} {\bibfnamefont {H.}~\bibnamefont
  {Ji}}, \bibinfo {author} {\bibfnamefont {B.}~\bibnamefont {Zhang}}, \bibinfo
  {author} {\bibfnamefont {W.}~\bibnamefont {Wang}}, \bibinfo {author}
  {\bibfnamefont {L.}~\bibnamefont {Lv}},\ and\ \bibinfo {author}
  {\bibfnamefont {J.}~\bibnamefont {Shen}},\ }\bibfield  {title} {\bibinfo
  {title} {Ultraviolet light-induced terahertz modulation of an indium oxide
  film},\ }\href {https://doi.org/10.1364/OE.26.007204} {\bibfield  {journal}
  {\bibinfo  {journal} {Opt. Express}\ }\textbf {\bibinfo {volume} {26}},\
  \bibinfo {pages} {7204} (\bibinfo {year} {2018})}\BibitemShut {NoStop}%
\bibitem [{\citenamefont {Simpkins}\ \emph {et~al.}(2021)\citenamefont
  {Simpkins}, \citenamefont {Dunkelberger},\ and\ \citenamefont
  {Owrutsky}}]{Simpkins2021}%
  \BibitemOpen
  \bibfield  {author} {\bibinfo {author} {\bibfnamefont {B.~S.}\ \bibnamefont
  {Simpkins}}, \bibinfo {author} {\bibfnamefont {A.~D.}\ \bibnamefont
  {Dunkelberger}},\ and\ \bibinfo {author} {\bibfnamefont {J.~C.}\ \bibnamefont
  {Owrutsky}},\ }\bibfield  {title} {\bibinfo {title} {Mode-specific chemistry
  through vibrational strong coupling (or a wish come true)},\ }\bibfield
  {booktitle} {\emph {\bibinfo {booktitle} {The Journal of Physical Chemistry
  C}},\ }\href {https://doi.org/10.1021/acs.jpcc.1c05362} {\bibfield  {journal}
  {\bibinfo  {journal} {The Journal of Physical Chemistry C}\ }\textbf
  {\bibinfo {volume} {125}},\ \bibinfo {pages} {19081} (\bibinfo {year}
  {2021})}\BibitemShut {NoStop}%
\end{thebibliography}%

\end{document}